\documentclass[pra,twocolumn,aps,amssymb,showpacs,superscriptaddress]{revtex4-1}

\usepackage{epsfig}
\usepackage{graphicx}
\usepackage{amsmath} 
\usepackage{amssymb}
\usepackage{enumerate}
\usepackage{hyperref}
\usepackage[normalem]{ulem}
\usepackage{cancel}

\usepackage{color}
\definecolor{rosso}{rgb}{1,0,0}
\definecolor{verde}{rgb}{0,1,0}
\definecolor{blue}{rgb}{0,0,1}
\definecolor{verdescuro}{rgb}{0,0.5,0.5}
\definecolor{rossoscuro}{rgb}{0.7,0.3,0}
\definecolor{bluscuro}{rgb}{0.3,0,0.7}
\definecolor{magenta}{rgb}{1,0,1}

\begin{document}

\title{Inclusion of pairing fluctuations in the differential equation for the gap parameter \\ for superfluid fermions in the presence of nontrivial spatial constraints}

\author{L. Pisani}
\affiliation{School of Science and Technology, Physics Division, Universit\`{a} di Camerino, 62032 Camerino (MC), Italy}
\author{V. Piselli}
\affiliation{School of Science and Technology, Physics Division, Universit\`{a} di Camerino, 62032 Camerino (MC), Italy}
\affiliation{CNR-INO, Istituto Nazionale di Ottica, Sede di Firenze, 50125 (FI), Italy}
\author{G. Calvanese Strinati}
\email{giancarlo.strinati@unicam.it}
\affiliation{School of Science and Technology, Physics Division, Universit\`{a} di Camerino, 62032 Camerino (MC), Italy}
\affiliation{CNR-INO, Istituto Nazionale di Ottica, Sede di Firenze, 50125 (FI), Italy}


\begin{abstract}
Most theoretical treatments of inhomogeneous superconductivity/fermionic superfluidity  have been based on the Bogoliubov-deGennes equations (or, else, on their various simplified forms), 
which implement a standard mean-field decoupling in the presence of spatial inhomogeneities.
This approach is reliable even at finite temperature for weak inter-particle attraction, when the Cooper pair size is much larger than the average inter-particle distance 
(corresponding to the BCS limit of the BCS-BEC crossover). 
However, it  looses accuracy for increasing attraction when the Cooper pair size becomes comparable or even smaller than the average inter-particle distance
(corresponding to the BEC limit of the BCS-BEC crossover), 
in particular when finite-temperature effects are considered.
In these cases, inclusion of pairing fluctuations beyond mean field is required, a task that  turns out to be especially difficult in the presence of inhomogeneities.
Here, we implement the inclusion of pairing fluctuations directly on a coarse-graining version of the Bogoliubov-deGennes equations, which makes it simpler and faster to obtain a solution over 
the whole sector of the temperature-coupling phase diagram of  the BCS-BEC crossover in the broken-symmetry phase.
We apply this method in the presence of a super-current flow, such that problems related to the Josephson effect throughout the BCS-BEC crossover can be addressed under a variety of circumstances.
This is relevant in the view of recent experimental data with ultra-cold Fermi atoms, to which the results of the present approach are shown to favorably compare in the companion article.
\end{abstract}

\maketitle

\section{Introduction} 
\label{sec:introduction}

Soon after the properties of \emph{homogeneous\/} fermionic superfluids have been accounted for on the basis of the BCS pairing approach \cite{BCS-1957,Parks-1969}, 
Gor'kov was able to extend this approach to deal with \emph{inhomogeneous\/} fermionic superfluids in terms of a many-body Green's functions formulation \cite{Gorkov-1958},
which specifically emphasizes that the resulting Gor'kov equations hold within a mean-field decoupling.
The goal of that approach was to accurately calculate physical properties that depend on the spatial profile of the gap (order) parameter $\Delta(\mathbf{r})$ in the presence of non-trivial geometrical constraints and/or confinements.
Later on, de Gennes reformulated the Gor'kov approach in terms of a complete set of fermionic single-particle wave functions in the superfluid phase, obtaining what are known as the Bogoliubov-deGennes (BdG) equations \cite{BdG-1966}.
The equivalence between the Gor'kov and de Gennes approaches can be demonstrated by expressing the Gor'kov single-particle Green's functions in terms of the de Gennes wave functions \cite{PS-2003}.

Numerical solutions of either Gor'kov or de Gennes approaches were mostly implemented in what is currently referred to as the weak-coupling (BCS) limit of the BCS-BEC crossover \cite{Physics-Reports-2018}, which is characterized by the presence of a well-defined underlying Fermi surface.
This presence has made it possible to introduce approximations to either the Gor'kov or the de Gennes approaches, that hold specifically in the BCS limit.
These include the derivation of the Ginzburg-Landau (GL) equation at temperatures close to the superfluid critical temperature $T_{c}$ \cite{Gorkov-1959}, as well as the Eilenberger \cite{Eilenberger-1968} and Usadel \cite{Usadel-1970} transport equations to deal with type-II and disordered superconductors, respectively.
These approximations, however, do not hold in the context of the BCS-BEC crossover, whereby the concept of an underlying sharp Fermi surface looses progressively its meaning away from the BCS limit.

In principle, the Gor'kov and de Gennes approaches can be applied to the whole BCS-BEC crossover, for given attractive inter-particle interaction and at any temperature below $T_{c}$, by taking care of the density equation 
that determines the evolution of the thermodynamic chemical potential throughout the crossover \cite{Physics-Reports-2018}.
This was done, for instance, for the Josephson effect at zero temperature in Ref.~\cite{SPS-2010} and for a single vortex at any temperature below $T_{c}$ in Ref.~\cite{SPS-2013}.
In this respect, one may recall what Leggett has recently emphasized \cite{Leggett-2017}, that the BdG equations have formed the \emph{basis\/} of almost all discussions of inhomogeneous superconductivity in the theoretical literature 
for the last fifty years. 

Nevertheless, efficient numerical methods to solve the BdG equations  in the presence of non-trivial geometrical constraints and/or confinements are still lacking.
This is because,  in spite of their apparent simplicity, the numerical solution of the BdG equations poses severe problems related both to computational time and memory space \cite{SPS-2010,SPS-2013}.
 These problems are related to the fact that  enforcing the Pauli principle requires detailed knowledge of a whole set of single-particle wave functions, 
even though one eventually ends up with the single function $\Delta(\mathbf{r})$ that accounts for the spatial dependence of the gap parameter.

To make these computational problems less severe, a (highly) non-linear \emph{differential\/} equation for the gap parameter $\Delta(\mathbf{r})$ (called the LPDA equation, for it entails a Local Phase Density Approximation) was introduced  in Ref.~\cite{SS-2014}, by  performing  a suitable double-coarse-graining procedure on the BdG equations 
 which affects the magnitude and phase of the gap parameter separately, and  through which an averaging over the above set of single-particle wave functions was effectively achieved.
The LPDA equation was also shown analytically to reduce to the Ginzburg-Landau (GL) equation for Cooper pairs in the weak-coupling (BCS) limit at temperatures close to $T_{c}$ \cite{Gorkov-1959}, as well as to the Gross-Pitaevskii (GP) equation for composite bosons in the strong-coupling (BEC) limit at low temperature \cite{PS-2003}.
The LPDA equation thus represents a suitable generalization of both GL and GP equations over an extended sector of the coupling-temperature phase diagram of the BCS-BEC crossover
(whereby, by tuning the inter-particle coupling, the system evolves from a BCS state with largely overlapping pairs of opposite-spin fermions that obey Fermi statistics, to a BEC state with dilute fermionic dimers that obey 
Bose statistics \cite{Physics-Reports-2018}).

In addition, owing to this capability of spanning the whole BCS-BEC crossover, the differential LPDA equation has a definite advantage over the mean-field Eilenberger \cite{Eilenberger-1968} and Usadel \cite{Usadel-1970} equations, which, as already mentioned, instead apply only in the weak-coupling (BCS) limit by assuming that the phenomenon of superconductivity occurs near the Fermi surface.

In practice, solving for the LPDA equation leads to a considerable reduction of time and storage requirements with respect to the BdG equations \cite{SS-2014}.
The LPDA approach was then utilized to account for the generation of complex vortex patterns in a trapped Fermi gas undergoing the BCS-BEC crossover \cite{SPS-2015}, and more recently was applied to investigate several phenomena related to a super-current flow in the context of the Josephson effect at finite temperature along the BCS-BEC crossover \cite{PSS-2020}.
It should also be mentioned that a non-local (integral) version of the LPDA equation was considered  in Ref.~\cite{SS-2017}, where it was shown that the length scale of the ``granularity'' resulting from the double-coarse-graining procedure, on which the LPDA approach rests, corresponds to the Cooper pair size at any coupling and temperature.
For this reason, the results obtained by solving the differential LPDA equation are expected to be most reliable when the profiles of (the magnitude and phase of) $\Delta(\mathbf{r})$ vary smoothly over a spatial range not smaller than this coarse-graining length scale.
The integral version of the LPDA equation was further utilized in Ref.~\cite{PSS-2018} in the context of the proximity effect occurring at the interface between two different fermionic superfluids.

Even though solving for the LPDA equation in the place of the BdG equations represents a useful practical improvement when addressing problems of physical interest 
across the BCS-BEC crossover,
 it is clear that the LPDA approach can at most recover the results obtained by solving the original BdG equations, but not more than that.
In particular, the mean-field decoupling for the inhomogeneous case (on which the BdG approach rests) is expected not to be appropriate \emph{at finite temperature\/}, when the inter-particle coupling is increased from the (BCS) weak-coupling to the (BEC) strong-coupling limit.
This is in line to what happens for the homogeneous case, when the mean-field decoupling alone is not able to recover the Bose-Einstein critical temperature in the BEC limit of the BCS-BEC crossover \cite{Physics-Reports-2018}.
As originally pointed out in Ref.~\cite{NSR-1985}, inclusion of \emph{pairing fluctuations\/} beyond mean field is thus required to obtain reliable results at finite temperature,
 and not only on the BEC side of the crossover but also in the unitary regime  intermediate between the BCS and BEC limits, where the Cooper pair size is comparable with the average inter-particle distance.

In principle, the inclusion of pairing fluctuations beyond mean field for the inhomogeneous case could be implemented by a suitable generalization of the BdG equations.
An initial attempt in this direction was made in Ref.~\cite{Kohn-1988} through an extension of the density-functional theory to superconductors, although this approach  was not fully applied in practice to specific problems.
A related approach was later developed within the so-called Superfluid  Local Density Approximation (SLDA), which was implemented for a limited number of inhomogeneous situations (and mostly at zero temperature) \cite{Bulgac-2012}.
Both these approaches retain the formal structure and the ensuing numerical complexity of the BdG equations, by solving for a system of coupled differential equations to obtain a whole set of two-component single-particle-like fermionic wave functions, which somewhat limits the feasibility for applying them in practice.

\begin{figure}[t]
\begin{center}
\includegraphics[width=8.9cm,angle=0]{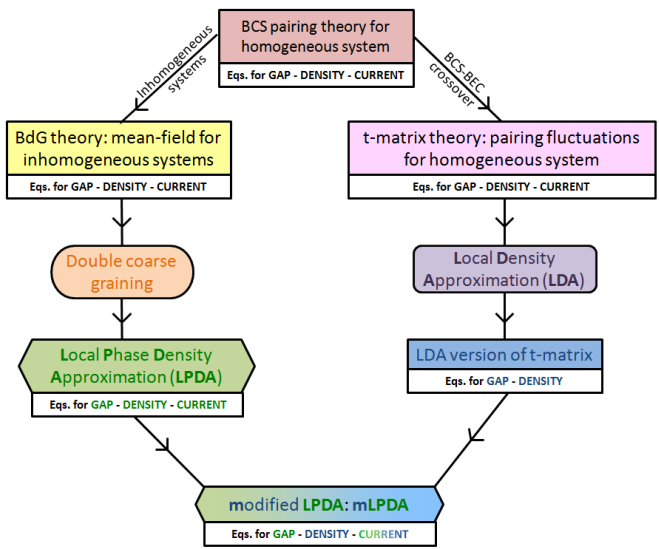}
\caption{Schematic ``flow diagram'' giving the prescription for obtaining the mLPDA approach. 
              At a first step, the LPDA equation is obtained by a suitable coarse graining of the BdG equations \cite{SS-2014}.
              At a second step, the LPDA equation is merged with a local version of the $t$-matrix approach of Ref.~\cite{PPS-2004}, yielding eventually the mLPDA approach.}
\label{Figure-1}
\end{center} 
\end{figure}  

In the spirit of the LPDA approach discussed above, we prefer to retain the advantages of solving a single (although highly non-linear) differential equation directly for the gap parameter $\Delta(\mathbf{r})$,
and to include the effects of pairing fluctuations for the inhomogeneous Fermi system of interest directly on top of the LPDA equation, which was itself derived from an inhomogeneous mean-field decoupling.
To this end, we take advantage of the approach of Ref.~\cite{PPS-2004}, where pairing fluctuations were added at the level of the $t$-matrix approximation 
in the broken-symmetry phase on top of the homogeneous mean-field approach.
Specifically, in Ref.~\cite{PPS-2004} the gap and density equations for the homogeneous Fermi system were treated on a different footing, in the sense that the gap equation was retained 
in the form valid at the mean-field level while the density equation was modified by including pairing fluctuations at the level of the $t$-matrix approximation in the superfluid phase.
Similarly, for the inhomogeneous Fermi system of interest here, we will maintain the LPDA equation for the gap parameter in the original form introduced in Ref.~\cite{SS-2014},
but at the same time we will modify the expression for the local particle density $n(\mathbf{r})$ (and the local particle current $\mathbf{j}(\mathbf{r})$, when needed) by the inclusion of pairing fluctuations in the spirit of a local-density approach. 
This replacement effectively transforms the LPDA approach of Ref.~\cite{PPS-2004} into the \emph{modified mLPDA approach\/} 
(where $m$ stands for ``modified''), to be discussed in detail below.
The flow diagram depicted in Fig.~\ref{Figure-1}  shows schematically how the LPDA and $t$-matrix approaches merge with each other, giving rise to the mLPDA approach.

For definiteness, in this article we implement the mLPDA approach by dealing with the Josephson effect at finite temperature throughout the BCS-BEC crossover, whereby 
pairing fluctuations need to be included in the expressions of both the local particle density $n(\mathbf{r})$ \emph{and\/} current $\mathbf{j}(\mathbf{r})$.
In this case, the requirement for the particle current to maintain everywhere a uniform value will act as a constraint on the numerical solution of the mLPDA equation, 
in close analogy to what was done in Ref.~\cite{SS-2014} for the solution of the LPDA equation without pairing fluctuations.
In this way, we will be able to monitor quantitatively the changes introduced by pairing fluctuations in physical quantities related to the Josephson effect (like the critical current), 
when comparing with the results of Ref.~\cite{PSS-2020} where $n(\mathbf{r})$ and $\mathbf{j}(\mathbf{r})$ were instead treated at the mean-field level. 
In this context,  a \emph{generalized\/}  two-fluid model at finite temperature will be introduced, which evolves from its fermionic 
(Bardeen-like \cite{Bardeen-1958})  version in the BCS limit to its bosonic (Landau-like \cite{Landau-1941})  version in the BEC limit of the BCS-BEC crossover.
In addition, the mLPDA approach will enable us to obtain a favorable comparison with the experimental data on the Josephson effect with ultra-cold Fermi gases, which are available 
from Ref.~\cite{Kwon-2020} at low temperature across the BCS-BEC crossover and from Ref.\cite{DelPace-2021} over an extended temperature range at unitarity.
A  detailed comparison in this respect  is reported in the companion article \cite{PPS-companion-PRL-article}.

The diagrammatic $t$-matrix approach for the superfluid phase, here considered in conjunction with the LPDA equation to yield the mLPDA approach for dealing with inhomogeneous spatial situations while spanning the BCS-BEC crossover, is an approximation that has often been adopted for including pairing fluctuations beyond mean field in the homogeneous case. 
And this was done not only in Ref.~\cite{PPS-2004}, on whose approach we specifically rely in the present article, but also in the context of the diagrammatic self-consistent version of Ref.~\cite{Haussmann-2007}
as well as of functional integrals \cite{HLD-2006} where the $t$-matrix is known as Gaussian approximation. 

In this article, we focus on the mLPDA approach as schematically summarized in  Fig.~\ref{Figure-1}, which we regard as a proof-of-principle for the way how pairing fluctuations and spatial inhomogeneities can be dealt with simultaneously also in the presence of a supercurrent.
In perspective, however, one may anticipate that the mLPDA approach is amenable to systematic improvements through a ``modular'' inclusion of additional many-body diagrammatic contributions, 
over and above those already considered by the $t$-matrix approach.
This inclusion can be conveniently organized by relying on the method introduced in Ref.~\cite{PPSb-2018}, where the gap equation even beyond mean field was cast in the alternative form of
a generalized Hugenholtz-Pines condition for fermion pairs, thereby extending to the whole BCS-BEC crossover the validity of this condition originally conceived for point-like bosons \cite{Hugenholtz-Pines-1959}.
In Ref.~\cite{PPSb-2018} this generalized Hugenholtz-Pines condition was considered for the homogeneous case only.
However, it is possible to adapt it to the presence of spatial inhomogeneities, by
(i) first replacing the homogeneous gap equation at the mean-field level with the LPDA equation of Ref.~\cite{SS-2014} where the kinetic energy required for spatial inhomogeneities is duly taken into account, 
and (ii) then adding to the LPDA equation the bosonic-like self-energy corrections introduced in Ref.~\cite{PPSb-2018} for the homogeneous case, but now treated within a local-density approximation to
account for spatial inhomogeneities.
For the needs of the BCS-BEC crossover, what has now become a \emph{local\/} Hugenholtz-Pines condition for fermion pairs has to be further supplied by suitable expressions for the particle and current densities.
Altogether, this appears to be a rather ambitious yet promising program, which for being implemented would require one to undertake considerable numerical efforts (especially in the presence of a supercurrent).
Full implementation of this program thus definitely remains outside the objectives of the present article, which focuses instead on the novelty of mLPDA approach itself.
Nevertheless, a preliminary albeit partial attempt to go beyond the mLPDA approach along these lines is considered in the companion article \cite{PPS-companion-PRL-article}, to test how it could improve
on the comparison with experimental data. 
In Ref.\cite{PPS-companion-PRL-article} the mLPDA approach is improved according to the scheme outline above, by considering the bosonic-like self-energy corrections of the Popov \cite{PS-2005} and Gorkov-Melik-Barkhudarov (GMB) \cite{GMB-1961} type in their extended versions established in Ref.~\cite{PPSb-2018}, without, however, including the effect of a supercurrent in the corresponding diagrammatic contributions.

The present article is organized as follows.
Section~\ref{sec:theoretical_approach} sets up the theoretical framework through which pairing fluctuations can be included in the LPDA approach, specifically in the context of the Josephson effect.
Section~\ref{sec:numerical_procedures} describes the numerical procedures that we have exploited in this context.
Section~\ref{sec:numerical_results} reports on a number of physical results for which pairing fluctuations play an essential role in the context of the Josephson effect.
Section~\ref{sec:conclusions} gives our conclusions.
Appendix~\ref{sec:Appendix-A} derives the expression of the particle-particle ladder in the presence of a super-current, 
 and Appendix~\ref{sec:Appendix-B} discusses how the bosonic two-fluid model is obtained analytically in the BEC limit of our expression for the current in the presence of pairing fluctuations.
Throughout, we shall consider balanced spin populations and set $\hbar = 1$ for convenience.

\section{Theoretical approach} 
\label{sec:theoretical_approach}

In this Section, we describe in detail the steps required to implement the inclusion of pairing fluctuations in the LPDA approach of Ref.~\cite{SS-2014}, 
thus yielding to what we refer to as the mLPDA approach, following the procedure that is schematically summarized in  Fig.~\ref{Figure-1}.

As mentioned in the Introduction, for definiteness we shall find it convenient to discuss the implementation from the LPDA to the mLPDA approaches in the context of the Josephson effect, which was recently studied  at the level 
of the LPDA equation as a function of both temperature and inter-particle attraction along the BCS-BEC crossover \cite{PSS-2020}.
Accordingly, the mLPDA approach will differ from the LPDA approach of Ref.~\cite{SS-2014} in the expressions of the \emph{local\/} particle density $n(\mathbf{r})$ \emph{and\/} current density $\mathbf{j}(\mathbf{r})$,
since these now include the effect of pairing fluctuations beyond mean field.
 At the same time, however, the differential LPDA equation of Ref.~\cite{SS-2014} will keep its formal structure.
In particular, pairing fluctuations will be introduced by exploiting the $t$-matrix approach of Ref.~\cite{PPS-2004}, which will now be reframed to take into account the presence of a supercurrent 
that affects in a consistent way the single-particle Green's functions entering the many-body diagrammatic expression of the $t$-matrix itself.
A  task of primary importance is thus finding suitable expressions for $n(\mathbf{r})$ and $\mathbf{j}(\mathbf{r})$, which  take into account the
effect of local pairing fluctuations.

Quite generally, the local number density and current in the broken-symmetry phase can be  expressed in the form
\begin{eqnarray}
n(\mathbf{r}) & = & \frac{2}{\beta} \sum_{n} e^{i \omega_{n} \eta} \, G_{11}(\mathbf{r},\mathbf{r};\omega_{n})
\label{local-density} \\
\mathbf{j}(\mathbf{r}) & = & \frac{1}{\beta} \sum_{n} e^{i \omega_{n} \eta} \frac{(\nabla_{\mathbf{r}} - \nabla_{\mathbf{r'}})}{i m} \, G_{11}(\mathbf{r},\mathbf{r'};\omega_{n})|_{\mathbf{r}=\mathbf{r'}} \, .
\label{local-current}
\end{eqnarray}
Here, $\beta = (k_{B} T)^{-1}$ is the inverse temperature ($k_{B}$ being the Boltzmann constant), $\eta$ a positive infinitesimal, $m$ the fermion mass, and $\omega_{n} = (2n+1)\pi/\beta$
($n$ integer) a fermionic Matsubara frequency \cite{Schrieffer-1964}.
The ``normal'' single-particle Green's function $G_{11}$ entering the above expressions can, in turn, be obtained together with its ``anomalous'' counterpart $G_{12}$, 
by solving the Dyson equation in the broken-symmetry phase \cite{Strinati-1998}
\begin{small}
\begin{eqnarray}
& - & \! \frac{\partial}{\partial \tau_{1}} G_{i_{1}i_{2}}(x_{1},x_{2}) + \sum_{i_{3}} \sigma^{(3)}_{i_{1}i_{3}} \!\! \left( \! \frac{\nabla_{1}^{2}}{2 m} - V_{\mathrm{ext}}(\mathbf{r}_{1}) + \mu \! \right) \! G_{i_{3}i_{2}}(x_{1},x_{2})
\nonumber \\
& - & \int \! dx_{3} \sum_{i_{3}} \Sigma_{i_{1}i_{3}}(x_{1},x_{3}) \, G_{i_{3}i_{2}}(x_{3},x_{2}) = \delta_{i_{1} i_{2}} \delta(x_{1}-x_{2}) \, ,
\label{Dyson-equation}
\end{eqnarray}
\end{small}
\noindent
where the imaginary time $\tau$ is limited by $0 \le \tau \le \beta$, $x=(\mathbf{r},\tau)$ is a four-variable, $i=(1,2)$ a Nambu index \cite{Nambu-1960}, $\sigma^{(3)}$ the third Pauli matrix, $V_{\mathrm{ext}}(\mathbf{r})$ an external (like a trapping) potential, $\mu$ the thermodynamic chemical potential, and $\Sigma_{ii'}(x,x')$ are elements of the self-energy.
At thermodynamic equilibrium, all quantities entering Eq.~(\ref{Dyson-equation}) depend only on the difference $\tau - \tau'$ of the imaginary-time variables.
In particular, at the mean-field level  \cite{Schrieffer-1964} one takes $\Sigma_{12}(x_{1},x_{2}) = - \delta(x_{1}^{+} - x_{2}) \Delta(\mathbf{r}_{1})$ for the off-diagonal component (where $x_{1}^{+}$ signifies that the time variable is augmented by a positive infinitesimal) and neglect the diagonal components of the self-energy.

In our case, we need an expression for $G_{11}(\mathbf{r},\mathbf{r'};\omega_{n})$ to be used in Eqs.~(\ref{local-density}) and (\ref{local-current}), which takes consistently into account the presence of a supercurrent flowing through the Fermi system of interest.
The remaining part of this Section is devoted to this purpose under different relevant conditions.

\begin{center}
{\bf A. Effects of a uniform flow on the Dyson \\ equation in the homogeneous case}
\end{center}

To construct the expressions for $n(\mathbf{r})$ and $\mathbf{j}(\mathbf{r})$ we are after, which include local pairing fluctuations in the context of the Josephson effect, it is convenient to 
first consider the homogeneous case when the external potential $V_{\mathrm{ext}}(\mathbf{r})$ in Eq.~(\ref{Dyson-equation}) is set to vanish.
In this case, when a uniform supercurrent flows through a homogeneous environment, the gap parameter of the superfluid Fermi system takes the form \cite{BdG-1966}
\begin{equation}
\Delta(\mathbf{r}) = e^{i 2 \mathbf{q} \cdot \mathbf{r}} \, \Delta_{\mathbf{q}} \, .
\label{gap-supercurrent-homogeneous}
\end{equation}
Here, $\mathbf{q}$ is a wave vector in the direction of the flow  that enters the linear phase $2 \mathbf{q} \! \cdot \! \mathbf{r}$ of the gap parameter, while $\Delta_{\mathbf{q}}$ stands for the 
associate magnitude of the gap parameter (which at zero temperature may itself depend on the magnitude of $\mathbf{q}$ when this exceeds a critical value \cite{BdG-1966}).

Quite generally, to comply with the  spatial dependence of $\Delta(\mathbf{r})$ of the form (\ref{gap-supercurrent-homogeneous}) 
(not only at the mean-field level but also beyond it), we proceed \emph{operatively\/} in the following way.
The single-particle Green's functions entering the Dyson equation (\ref{Dyson-equation}) are taken of the form \cite{footnote-local-phase-transformation}
\begin{eqnarray}
G_{11}(x,x';\mathbf{q}) & = & e^{i \mathbf{q} \cdot (\mathbf{r} - \mathbf{r'})} \, \mathcal{G}_{11}(x-x';\mathbf{q})
\label{G11-and-q-vector} \\
G_{12}(x,x';\mathbf{q}) & = & e^{i \mathbf{q} \cdot (\mathbf{r} + \mathbf{r'})} \, \mathcal{G}_{12}(x-x';\mathbf{q})
\label{G12-and-q-vector} \\
G_{21}(x,x';\mathbf{q}) & = & e^{- i \mathbf{q} \cdot (\mathbf{r} + \mathbf{r'})} \, \mathcal{G}_{21}(x-x';\mathbf{q})
\label{G21-and-q-vector} \\
G_{22}(x,x';\mathbf{q}) & = & e^{- i \mathbf{q} \cdot (\mathbf{r} - \mathbf{r'})} \, \mathcal{G}_{22}(x-x';\mathbf{q})
\label{G22-and-q-vector}
\end{eqnarray}
where
\begin{equation}
\mathcal{G}_{ii'}(x-x';\mathbf{q}) = \sum_{k} e^{i \mathbf{k} \cdot (\mathbf{r} - \mathbf{r'})} e^{-i \omega_{n} (\tau - \tau')} \, \mathcal{G}_{ii'}(k;\mathbf{q}) \, ,
\label{Gij-Fourier-transform}
\end{equation}
with the fermionic four-vector $k = (\mathbf{k},\omega_{n})$ and the short-hand notation
\begin{equation}
\sum_{k} \, \longleftrightarrow \, \int \!\!\! \frac{d \mathbf{k}}{(2 \pi)^{3}} \, \frac{1}{\beta} \sum_{n} \, .
\label{notation-integral-sum-fermionic}
\end{equation}

Note that in Eqs.~(\ref{G11-and-q-vector})-(\ref{G22-and-q-vector}) the diagonal elements depend on $\mathbf{r} - \mathbf{r'}$ while the off-diagonal elements depend on $\mathbf{r} + \mathbf{r'}$.
This is due to the way the field operators and their adjoints enter the expressions for the single-particle Green's functions $G_{ij}$ in the broken-symmetry phase \cite{Schrieffer-1964}.

We further assume that the self-energies $\Sigma_{ii'}(x,x')$ in the Dyson equation (\ref{Dyson-equation}) share the same dependence on $\mathbf{q}$ of the expressions 
(\ref{G11-and-q-vector})-(\ref{G22-and-q-vector}).
[This property  will be explicitly verified in Appendix~\ref{sec:Appendix-A} within the $t$-matrix approximation.]
In this way, the convolutions entering the last term on  the left-hand side of the Dyson equation (\ref{Dyson-equation}) considerably simplify.
For instance,
\begin{eqnarray}
&& \int \! dx_{3} \, \Sigma_{11}(x_{1},x_{3}) \, G_{11}(x_{3},x_{2}) \hspace{0.5cm} \longleftrightarrow  
\nonumber \\ 
&& e^{i \mathbf{q} \cdot (\mathbf{r}_{1} - \mathbf{r}_{2})} \! \int \! dx_{3} \, \mathfrak{S}_{11}(x_{1} - x_{3};\mathbf{q}) \, \mathcal{G}_{11}(x_{3} - x_{2};\mathbf{q})
\label{convolution-11-11}
\end{eqnarray}
and 
\begin{eqnarray}
&& \int \! dx_{3} \, \Sigma_{12}(x_{1},x_{3}) \, G_{21}(x_{3},x_{2}) \hspace{0.5cm} \longleftrightarrow 
\nonumber \\ 
&& e^{i \mathbf{q} \cdot (\mathbf{r}_{1} - \mathbf{r}_{2})} \! \int \! dx_{3} \, \mathfrak{S}_{12}(x_{1} - x_{3};\mathbf{q}) \, \mathcal{G}_{21}(x_{3} - x_{2};\mathbf{q}) \, .
\label{convolution-12-21}
\end{eqnarray}

Note that, while the symbol $\Sigma_{ij}$ for the self-energy is associated with the full single-particle Green's function $G_{ij}$ like in Eq.~(\ref{Dyson-equation}), the symbol $\mathfrak{S}_{ij}$ 
for the \emph{reduced\/} self-energy is instead associated with the reduced single-particle Green's function $\mathcal{G}_{ij}$, as defined in Eqs.~(\ref{G11-and-q-vector})-(\ref{G22-and-q-vector}) 
once all factors $e^{\pm i \mathbf{q} \cdot (\mathbf{r} \pm \mathbf{r'})}$ are conveniently disposed off.

By introducing also the expansion
\begin{equation}
\mathfrak{S}_{ii'}(x-x';\mathbf{q}) = \sum_{k} e^{i \mathbf{k} \cdot (\mathbf{r} - \mathbf{r'})} e^{-i \omega_{n} (\tau - \tau')} \, \mathfrak{S}_{ii'}(k;\mathbf{q})
\label{Sigmaij-Fourier-transform}
\end{equation}
in analogy with Eq.~(\ref{Gij-Fourier-transform}), the convolutions like those on  the right-hand sides of Eqs.~(\ref{convolution-11-11}) and (\ref{convolution-12-21}) become:
\begin{widetext}
\begin{equation}
\int \! dx_{3} \, \mathfrak{S}_{i_{1}i_{3}}(x_{1} - x_{3};\mathbf{q}) \, \mathcal{G}_{i_{3}i_{2}}(x_{3} - x_{2};\mathbf{q}) 
= \sum_{k} e^{i \mathbf{k} \cdot (\mathbf{r}_{1} - \mathbf{r}_{2})} \, e^{-i \omega_{n} (\tau_{1} - \tau_{2})} \, \mathfrak{S}_{i_{1}i_{3}}(k;\mathbf{q}) \,\, \mathcal{G}_{i_{3}i_{2}}(k;\mathbf{q}) \, .
\label{convolution-generic} 
\end{equation}
With the further result
\begin{equation}
\nabla^{2} \left( e^{\pm i \mathbf{q} \cdot \mathbf{r}} \mathcal{G}_{ii'}(x - x';\mathbf{q}) \right)
= - e^{\pm i \mathbf{q} \cdot \mathbf{r}} \sum_{k} e^{i \mathbf{k} \cdot (\mathbf{r} - \mathbf{r'})} e^{-i \omega_{n} (\tau - \tau')} \, (\mathbf{k} \pm \mathbf{q})^{2} \, \mathcal{G}_{ii'}(k;\mathbf{q}) \, ,
\label{nabla-square} 
\end{equation}
the Dyson equation (\ref{Dyson-equation}) reduces eventually to a set of four coupled algebraic equations, that can conveniently be cast  in matrix form
\begin{equation}
\left[ \! \begin{array}{cc} i \omega_{n} - \xi(\mathbf{k}+\mathbf{q}) - \mathfrak{S}_{11}(k;\mathbf{q}) & - \mathfrak{S}_{12}(k;\mathbf{q}) \\
                                          - \mathfrak{S}_{21}(k;\mathbf{q}) & i \omega_{n} + \xi(\mathbf{k}-\mathbf{q}) - \mathfrak{S}_{22}(k;\mathbf{q}) \end{array} \! \right] \,
\left[ \begin{array}{cc} \mathcal{G}_{11}(k;\mathbf{q}) & \mathcal{G}_{12}(k;\mathbf{q}) \\
                                                      \mathcal{G}_{21}(k;\mathbf{q}) & \mathcal{G}_{22}(k;\mathbf{q}) \end{array} \right]
= \left[ \begin{array}{cc} 1 & 0 \\ 0 & 1 \end{array} \right]
\label{algebraic-Dyson-equation}  
\end{equation}
with the notation $\xi(\mathbf{k}) = \mathbf{k}^{2}/(2m) - \mu$.
Solution to this matrix equation yields:
\begin{eqnarray}
\left[ \! \begin{array}{cc} \mathcal{G}_{11}(k;\mathbf{q}) & \mathcal{G}_{12}(k;\mathbf{q}) \\ \mathcal{G}_{21}(k;\mathbf{q}) & \mathcal{G}_{22}(k;\mathbf{q}) \end{array} \! \right]
& = & \frac{1}{ [i \omega_{n} - \xi(\mathbf{k}+\mathbf{q}) - \mathfrak{S}_{11}(k;\mathbf{q})] \,\, [i \omega_{n} + \xi(\mathbf{k}-\mathbf{q}) - \mathfrak{S}_{22}(k;\mathbf{q})] 
                       -  \mathfrak{S}_{12}(k;\mathbf{q}) \mathfrak{S}_{21}(k;\mathbf{q})} 
\nonumber \\
& \times &  \left[ \! \begin{array}{cc} i \omega_{n} + \xi(\mathbf{k}-\mathbf{q}) - \mathfrak{S}_{22}(k;\mathbf{q}) & \mathfrak{S}_{12}(k;\mathbf{q}) \\
                                                       \mathfrak{S}_{21}(k;\mathbf{q}) & i \omega_{n} - \xi(\mathbf{k}+\mathbf{q}) - \mathfrak{S}_{11}(k;\mathbf{q}) \end{array} \! \right]
\label{algebraic-Dyson-equation-solution}
\end{eqnarray}
\end{widetext}
where the properties $\mathfrak{S}_{22}(k;\mathbf{q}) = - \mathfrak{S}_{11}(-k;\mathbf{q})$ and $\mathfrak{S}_{12}(k;\mathbf{q}) = \mathfrak{S}_{21}(k;\mathbf{q})$ hold 
(analogous relations hold for the components of $\mathcal{G}_{ii'}(k;\mathbf{q})$).

The mean-field and $t$-matrix approximations correspond to different choices for the reduced self-energy $\mathfrak{S}_{ij}$, which we are now going to consider separately.

\begin{center}
{\bf B. Results for the homogeneous case \\ at the mean-field level}
\end{center}

We first specify the general results of Sec.~\ref{sec:theoretical_approach}-A to the mean-field (mf) case.
In this case, we take  $\mathfrak{S}_{11}^{\mathrm{mf}}(k;\mathbf{q}) = \mathfrak{S}_{22}^{\mathrm{mf}}(k;\mathbf{q}) = 0$ 
and $\mathfrak{S}_{12}^{\mathrm{mf}}(k;\mathbf{q}) = \mathfrak{S}_{21}^{\mathrm{mf}}(k;\mathbf{q}) = - \Delta_{\mathbf{q}}$,
where $\Delta_{\mathbf{q}}$ is the quantity entering Eq.~(\ref{gap-supercurrent-homogeneous}).
Accordingly, the solutions (\ref{algebraic-Dyson-equation-solution}) acquire the form:
\begin{eqnarray}
\mathcal{G}_{11}^{\mathrm{mf}}(k;\mathbf{q}) & = & \frac{u(\mathbf{k};\mathbf{q})^{2}}{i \omega_{n} - E_{+}(\mathbf{k};\mathbf{q})} + \frac{v(\mathbf{k};\mathbf{q})^{2}}{i \omega_{n} + E_{-}(\mathbf{k};\mathbf{q})}
\label{simplified-solutions-mean-field-11} \\
\mathcal{G}_{12}^{\mathrm{mf}}(k;\mathbf{q}) & = & - \, u(\mathbf{k};\mathbf{q}) \, v(\mathbf{k};\mathbf{q}) \, \left[ \frac{1}{i \omega_{n} - E_{+}(\mathbf{k};\mathbf{q})} \right.
\nonumber \\
& - & \left. \frac{1}{i \omega_{n} + E_{-}(\mathbf{k};\mathbf{q})} \right]
\label{simplified-solutions-mean-field-12} 
\end{eqnarray}
where
\begin{eqnarray}
u(\mathbf{k};\mathbf{q})^{2} = \frac{1}{2} \left( 1 + \frac{\xi(\mathbf{k};\mathbf{q})}{E(\mathbf{k};\mathbf{q})} \right)
\label{u-square} \\
v(\mathbf{k};\mathbf{q})^{2} = \frac{1}{2} \left( 1 - \frac{\xi(\mathbf{k};\mathbf{q})}{E(\mathbf{k};\mathbf{q})} \right)
\label{u-square}
\end{eqnarray}
with the notation
\begin{eqnarray}
\xi(\mathbf{k};\mathbf{q}) & = & \frac{\mathbf{k}^{2}}{2m} - \mu + \frac{\mathbf{q}^{2}}{2m}
\label{notation-I} \\
E(\mathbf{k};\mathbf{q}) & = & \sqrt{\xi(\mathbf{k};\mathbf{q})^{2} + \Delta_{\mathbf{q}}^{2}}
\label{notation-II} \\
E_{\pm} (\mathbf{k};\mathbf{q}) & = & E(\mathbf{k};\mathbf{q}) \pm \frac{\mathbf{k} \cdot \mathbf{q}}{m} \, .
\label{notation-III}
\end{eqnarray}
In these expressions, the wave vector $\mathbf{q}$ associated with the superfluid flow plays the role of an external parameter.
When $\mathbf{q}=0$, the expressions (\ref{simplified-solutions-mean-field-11})-(\ref{notation-III}) recover the standard results of the BCS theory \cite{FW-1971}.

For finite $\mathbf{q}$, the expressions (\ref{simplified-solutions-mean-field-11})-(\ref{notation-III}) have already been utilized in the context of the LPDA approach \cite{SS-2014} through a suitable extension to the inhomogeneous case (to be recalled below).

\vspace{0.5cm}
\begin{center}
{\bf C. Results for the homogeneous case \\ with the inclusion of pairing fluctuations }
\end{center}

As already anticipated, to include  in the results of Sec.~\ref{sec:theoretical_approach}-A the effect of pairing fluctuations over and above mean field, we shall rely 
on the $t$-matrix approach of Ref.~\cite{PPS-2004}. 
This has, however, to be suitably modified to account for the presence of a stationary superfluid flow in a consistent way.

In Appendix~\ref{sec:Appendix-A}, a detailed analysis is provided of the diagrammatic structure for the series of ladder diagrams in the broken-symmetry phase \emph{in the presence of a superfluid flow\/}.
This analysis  yields the following expressions for the reduced self-energies $\mathfrak{S}_{ii'}^{\mathrm{pf}}(k;\mathbf{q})$ with the inclusion of  pairing fluctuations (pf):
\begin{equation}
\hspace{-0.25cm} \left\{ 
\begin{array}{ll}
\!\! \mathfrak{S}_{11}^{\mathrm{pf}}(k;\mathbf{q}) \! = \! - \mathfrak{S}_{22}^{\mathrm{pf}}(-k;\mathbf{q}) \! = \! - \sum_{Q} \Gamma_{11}(Q;\mathbf{q}) \, \mathcal{G}_{11}^{\mathrm{mf}}(Q \! - \! k;\mathbf{q})   \\
\\
\!\! \mathfrak{S}_{12}^{\mathrm{pf}}(k;\mathbf{q}) \! = \! \mathfrak{S}_{21}^{\mathrm{pf}}(k;\mathbf{q}) = - \Delta_{\mathbf{q}} \, . 
\end{array}
\right.
\label{self-energies-t-matrix}
\end{equation}
Here, $Q = (\mathbf{Q},\Omega_{\nu})$ is a four-vector where $\Omega_{\nu} =  2\nu\pi/\beta$ ($\nu$ integer) is a bosonic Matsubara frequency, 
\begin{equation}
\sum_{Q} \, \longleftrightarrow \, \int \!\!\! \frac{d \mathbf{Q}}{(2 \pi)^{3}} \, \frac{1}{\beta} \sum_{\nu} 
\label{notation-integral-sum-bosonic}
\end{equation}
is a short-hand notation, the diagonal (normal) reduced single-particle Green's function $\mathcal{G}_{11}^{\mathrm{mf}}(k;\mathbf{q})$ is given by the mean-field expression (\ref{simplified-solutions-mean-field-11}), 
and the elements of the particle-particle ladder $\Gamma_{ii'}(Q;\mathbf{q})$ in the broken-symmetry phase are given by the expressions (\ref{full-particle-particle-ladder})-(\ref{B}) of Appendix~\ref{sec:Appendix-A},
(where they are also represented graphically in Fig.~\ref{Figure-8}  therein).

Entering the expressions (\ref{self-energies-t-matrix}) in Eq.~(\ref{algebraic-Dyson-equation-solution}) one obtains for the diagonal (normal) single-particle Green's function 
$\mathcal{G}_{11}^{\mathrm{pf}}(k;\mathbf{q})$ which includes pairing fluctuations in the presence of a superfluid flow: 
\begin{widetext}
\begin{large}
\begin{equation}
\mathcal{G}_{11}^{\mathrm{pf}}(k;\mathbf{q}) = 
\frac{1}{ i \omega_{n} - \xi(\mathbf{k}+\mathbf{q}) - \mathfrak{S}_{11}^{\mathrm{pf}}(k;\mathbf{q}) - \frac{\Delta_{\mathbf{q}}^{2}}{i \omega_{n} + \xi(\mathbf{k}-\mathbf{q}) + \mathfrak{S}_{11}^{\mathrm{pf}}(-k;\mathbf{q})}} \, .
\label{G-11-pairing-flucuations_and_current}
\end{equation}
\end{large}
\end{widetext}

Note that the mean-field result (\ref{simplified-solutions-mean-field-11}) can formally be obtained from the expression (\ref{G-11-pairing-flucuations_and_current}) by setting
$\mathfrak{S}_{11}^{\mathrm{pf}} = 0$ therein.

Next, we proceed to manipulate the expression (\ref{simplified-solutions-mean-field-11}) at the mean-field level and the expression (\ref{G-11-pairing-flucuations_and_current}) with the inclusion of pairing fluctuations, in order to extend them to the inhomogeneous case and  obtain the required expressions of the \emph{local\/} particle density and current.
These expressions will then be utilized, respectively, at the mean-field level thus recovering the LPDA approach of Ref.~\cite{SS-2014}, and with the inclusion of pairing fluctuations for implementing its mLPDA beyond-mean-field extension, which  constitutes the main novelty of the present article.

\begin{center}
{\bf D. Results for the inhomogeneous case \\ needed when solving the LPDA equation}
\end{center}  

The LPDA and mLPDA approaches differ from each other as far as the expressions (\ref{local-density}) of the local particle density and (\ref{local-current}) of the local particle current are concerned, 
in the sense that different forms of the single-particle Green’s function $G_{11}$ are utilized  in the two cases.

In the inhomogeneous case of interest, the single-particle Green’s function $G_{11}$ should, in principle, be obtained from the solution of the Dyson equation (\ref{Dyson-equation}) with an appropriate choice of the self-energy plus a given form of the  external potential $V_{\mathrm{ext}}(\mathbf{r})$, which is the source of inhomogeneities in the system.
In practice, however,  it is convenient to deal with this problem in a simplified way, by adopting a \emph{local perspective\/} that eventually leads to the 
LPDA and mLPDA approaches.

Although this strategy was already utilized in Ref.~\cite{SS-2014} in the context of the LPDA approach, it is convenient to briefly recall it here with the purpose of extending it to the mLPDA approach,
which depends on a different choice of the single-particle self-energy.

Quite generally, in a local approach one is willing  to approximate (cf. Eqs.~(\ref{G11-and-q-vector}) and (\ref{Gij-Fourier-transform}))
\begin{eqnarray}
& & G_{11}(x,x';\mathbf{q}) \longrightarrow
\label{G_11-local-approach} \\
& & \sum_{k} e^{i (\mathbf{k} + \mathbf{q}) \cdot (\mathbf{r} - \mathbf{r'})} e^{-i \omega_{n} (\tau - \tau')} \, \mathcal{G}_{11}(k;\mathbf{q}|(\mathbf{r}+\mathbf{r'})/2) 
\nonumber 
\end{eqnarray}
where the local version of $\mathcal{G}_{11}(k;\mathbf{q})$ needs to be suitably specified.
[The ``mid-point'' rule in Eq.~(\ref{G_11-local-approach}) is useful for the calculation of the current.]
At the level of the LPDA approach, one considers the mean-field expression for $\mathcal{G}_{11}^{\mathrm{mf}}(k;\mathbf{q})$ as given by Eq.~(\ref{simplified-solutions-mean-field-11}) 
and performs therein  the local replacements:
\begin{eqnarray}
\mu \, & \longrightarrow & \, \mu - V_{\mathrm{ext}}(\mathbf{r})
\label{replacement-mu} \\ 
\Delta_{\mathbf{q}} \, & \longrightarrow & \, \left| \Delta (\mathbf{r}) \right|
\label{replacement-magnitude-Delta} \\
\mathbf{q} \, & \longrightarrow & \, \mathbf{Q_{0}} + \nabla \phi(\mathbf{r}) \, .
\label{replacement-phase-Delta} 
\end{eqnarray}
Specifically, the replacements (\ref{replacement-magnitude-Delta}) and (\ref{replacement-phase-Delta}) are consistent with a gap parameter of the form 
$\Delta(\mathbf{r}) = |\Delta(\mathbf{r})| \exp\{i 2 [\mathbf{Q_{0}} \cdot \mathbf{r} + \phi(\mathbf{r})] \}$, where the wave vector $\mathbf{Q_{0}}$ would be  associated with a superfluid flow also in the
homogeneous case, while the additional phase $\phi(\mathbf{r})$ arises from the presence of the external potential $V_{\mathrm{ext}}(\mathbf{r})$
(which for the Josephson effect may correspond to a one-dimensional barrier embedded in an otherwise homogeneous superfluid - cf. Sec.~\ref{sec:numerical_procedures}).
In this way, the expression (\ref{local-density}) for the local particle density becomes
\begin{equation}
n(\mathbf{r}) = \int \!\!\! \frac{d \mathbf{k}}{(2 \pi)^{3}} \left\{ 1 - \frac{\xi(\mathbf{k};\mathbf{q}|\mathbf{r})}{E(\mathbf{k};\mathbf{q}|\mathbf{r})} \, 
[ 1 - 2 \, f_{F}(E_{+}(\mathbf{k};\mathbf{q}|\mathbf{r})) ] \right\}
\label{local-density-mean-field}
\end{equation}
where $f_{F}(\epsilon) = \left(e^{\beta \epsilon} + 1\right)^{-1}$ is the Fermi function and 
\begin{eqnarray}
& & \xi(\mathbf{k};\mathbf{q}|\mathbf{r}) = \frac{\mathbf{k}^{2}}{2m} - \mu + V_{\mathrm{ext}}(\mathbf{r}) + \frac{ \left( \mathbf{Q_{0}} + \nabla \phi(\mathbf{r}) \right)^{2}}{2m}
\label{notation-I-inhomogeneous} \\
& & E(\mathbf{k};\mathbf{q}|\mathbf{r}) = \sqrt{\xi(\mathbf{k};\mathbf{q}|\mathbf{r})^{2} + |\Delta(\mathbf{r})|^{2}}
\label{notation-II-inhomogeneous} \\
& & E_{+} (\mathbf{k};\mathbf{q}|\mathbf{r}) = E(\mathbf{k};\mathbf{q}|\mathbf{r}) + \frac{\mathbf{k}}{m} \cdot  \left( \mathbf{Q_{0}} + \nabla \phi(\mathbf{r}) \right) \, .
\label{notation-III-inhomogeneous}
\end{eqnarray}
are obtained from the expressions (\ref{notation-I})-(\ref{notation-III}) with the replacements (\ref{replacement-mu})-(\ref{replacement-phase-Delta}).

Similarly, for the local particle current one gets:
\begin{equation} 
\mathbf{j}(\mathbf{r}) = \frac{1}{m}  \left( \mathbf{Q_{0}} + \nabla \phi(\mathbf{r}) \right) n(\mathbf{r}) + 2 \! \int \!\!\! \frac{d \mathbf{k}}{(2 \pi)^{3}} \frac{\mathbf{k}}{m} f_{F}(E_{+}(\mathbf{k};\mathbf{q}|\mathbf{r}))
\label{local-current-mean-field}
\end{equation}
with $n(\mathbf{r})$ given by Eq.~(\ref{local-density-mean-field}).
The expression (\ref{local-current-mean-field}) was originally derived in Ref.~\cite{SS-2014}  where it was  tested for an isolated vortex, 
 was further utilized in Ref.~\cite{SPS-2015} for a complex array of vortices in a rotating trapped superfluid Fermi gas, and was finally  employed in Ref.~\cite{PSS-2020} 
in the context of the Josephson effect.
This expression generalizes to the inhomogeneous case the Bardeen  \emph{fermionic two-fluid model at finite temperature} \cite{Bardeen-1958}.

\begin{center}
{\bf E. Results for the inhomogeneous case \\ needed when solving the mLPDA equation}
\end{center}

To go beyond the mean-field level and include the effect of pairing fluctuations in the expressions of the local density and current needed for implementing the mLPDA approach, 
we have to convert the expression (\ref{G-11-pairing-flucuations_and_current}) for $\mathcal{G}_{11}^{\mathrm{pf}}(k;\mathbf{q})$ into a local version  
$\mathcal{G}_{11}^{\mathrm{pf}}(k;\mathbf{q}|\mathbf{r})$ that depends on the spatial coordinate $\mathbf{r}$.
To this end, we still rely on a local perspective as we did in Sec.~\ref{sec:theoretical_approach}-D for the single-particle Green's functions at the mean-field level, but we have now to extend it to
the particle-particle ladder that accounts for pairing fluctuations.

In this case, however, one cannot simply rely on making the local replacements (\ref{replacement-mu})-(\ref{replacement-phase-Delta}) whenever these quantities appear
in the expression (\ref{G-11-pairing-flucuations_and_current}).
This is because the self-energy $\mathfrak{S}_{11}^{\mathrm{pf}}(k;\mathbf{q})$ given by Eq.~(\ref{self-energies-t-matrix}) that enters the expression (\ref{G-11-pairing-flucuations_and_current}), in turn, contains
the diagonal element $\Gamma_{11}(Q;\mathbf{q})$ of the particle-particle ladder in the broken-symmetry phase given by Eqs.~(\ref{full-particle-particle-ladder})-(\ref{B}).
In those expressions, the gap equation (\ref{gap-equation-with_current}), relating the gap $\Delta_{\mathbf{q}}$ to the thermodynamic chemical potential $\mu$, guarantees the gapless condition at $Q=0$ for the particle-particle ladder in the presence of a current.
When taking into account (again, within \emph{a local perspective\/}) spatial inhomogeneities associated with  the presence of an external potential $V_{\mathrm{ext}}(\mathbf{r})$, 
one has thus to consider a local expression not only for the (magnitude and phase of the) gap parameter like in Eqs.~(\ref{replacement-magnitude-Delta}) and (\ref{replacement-phase-Delta}), but also for the chemical potential, in such a way to preserve \emph{at each spatial point\/} $\mathbf{r}$ the gapless condition at $Q=0$ for the desired local expression $\Gamma_{11}(Q;\mathbf{q}|\mathbf{r})$.
Preserving locally the gapless condition is, in fact, required  to avoid the occurrence of unphysical singularities in the particle-particle ladder everywhere in the  spatial regions occupied by the system (cf. Appendix~\ref{sec:Appendix-A}).
In practice, to make the above requirement satisfied, we will end up with associating an \emph{effective potential\/} $V_{\mathrm{eff}}(\mathbf{r})$ to a given external potential 
$V_{\mathrm{ext}}(\mathbf{r})$.

To achieve these goals within contained numerical efforts, we have adopted the following strategy: 

\vspace{0.1cm}
\noindent
(i) We have replaced $\Delta_{\mathbf{q}} \rightarrow |\Delta(\mathbf{r})|$ and $\mu \rightarrow \mu - V_{\mathrm{eff}}(\mathbf{r})$ in the quantities $\xi(\mathbf{k};\mathbf{q})$ 
and $E(\mathbf{k};\mathbf{q})$ given by Eqs.~(\ref{notation-I}) and (\ref{notation-II}) which enter the expressions (\ref{diagonal-bubble}) and (\ref{off-diagonal-bubble}) for the
particle-particle bubbles, where now $\mathbf{q} \rightarrow \mathbf{Q_{0}}$ without including the additional term $\nabla \phi(\mathbf{r})$ of Eq.~(\ref{replacement-phase-Delta}).
This is in the spirit of a Local Density Approximation (LDA) approach to $\Gamma_{11}(Q;\mathbf{Q_{0}}|\mathbf{r})$, whereby no gradient (either of the magnitude or the phase of the gap parameter) is taken into account.
We anticipate that the profile of $|\Delta(\mathbf{r})|$ is here provided by the solution of the differential mLPDA equation described in Sec.~\ref{sec:numerical_procedures} below which explicitly depends on the
external potential $V_{\mathrm{ext}}(\mathbf{r})$, in such a way that the following steps (ii)-(iv) will have to be repeated until self-consistency is achieved.

\vspace{0.1cm}
\noindent
(ii) We have  generated the shape  of the \emph{effective potential\/} $V_{\mathrm{eff}}(\mathbf{r})$  corresponding to a given  local gap profile $|\Delta(\mathbf{r})|$, 
by solving at each $\mathbf{r}$ the now local gap equation (\ref{gap-equation-with_current}) with $Q=0$, where the local replacements of step (i)  are also made 
\cite{footnote-opposite-procedure,PPPS-2004}.
With these values of $|\Delta(\mathbf{r})|$ and $V_{\mathrm{eff}}(\mathbf{r})$, we have then obtained the local diagonal element $\Gamma_{11}(Q;\mathbf{Q_{0}}|\mathbf{r})$ 
for given  $Q$, that results from the expressions given in Appendix~\ref{sec:Appendix-A} with the local forms of the reduced single-particle Green’s functions
(\ref{simplified-solutions-mean-field-11}) and (\ref{simplified-solutions-mean-field-12}).

\vspace{0.1cm}
\noindent
(iii) We have consistently made the above local replacements also in the single-particle Green's function $\mathcal{G}_{11}^{\mathrm{mf}}(Q \! - \! k;\mathbf{q})$ which closes the loop in the diagonal
self-energy (\ref{self-energies-t-matrix}), to eventually obtain the required local expression $\mathfrak{S}_{11}^{\mathrm{pf}}(k;\mathbf{Q_{0}}|\mathbf{r})$.

\vspace{0.1cm}
\noindent
(iv) We have finally obtained the expression for $\mathcal{G}_{11}^{\mathrm{pf}}(k;\mathbf{q}|\mathbf{r})$ by replacing in Eq.~(\ref{G-11-pairing-flucuations_and_current}) 
$\Delta_{\mathbf{q}} \rightarrow |\Delta(\mathbf{r})|$ and $\mathfrak{S}_{11}^{\mathrm{pf}}(k;\mathbf{q}) \rightarrow \mathfrak{S}_{11}^{\mathrm{pf}}(k;\mathbf{Q_{0}}|\mathbf{r})$, while in addition
\begin{eqnarray}
\xi(\mathbf{k \pm q}) & \rightarrow & \frac{\mathbf{k}^{2}}{2m} - \mu + V_{\mathrm{eff}}(\mathbf{r}) + \frac{ \left( \mathbf{Q_{0}} + \nabla \phi(\mathbf{r}) \right)^{2}}{2m}
\nonumber \\
& \pm & \frac{\mathbf{k}}{m} \cdot \left( \mathbf{Q_{0}} + \nabla \phi(\mathbf{r}) \right) \, .
\label{final-replacement}
\end{eqnarray}
Note that in Eq.~(\ref{final-replacement}) we have kept the full replacement (\ref{replacement-phase-Delta}) which includes $\nabla \phi(\mathbf{r})$, in such a way that the expression of 
$\mathcal{G}_{11}^{\mathrm{pf}}(k;\mathbf{q}|\mathbf{r})$ recovers that of $\mathcal{G}_{11}^{\mathrm{mf}}(k;\mathbf{q}|\mathbf{r})$ when the effect of pairing fluctuations is altogether neglected
(in this case the effective potential $V_{\mathrm{eff}}(\mathbf{r})$ would  consistently coincide with the external potential $V_{\mathrm{ext}}(\mathbf{r})$).
Once again, the profile $\phi(\mathbf{r})$ needed in Eq.~(\ref{final-replacement}) is consistently provided by the solution of the differential mLPDA equation described in Sec.~\ref{sec:numerical_procedures} below.

\vspace{0.1cm}
\noindent
(v) We have eventually approximated the expressions (\ref{local-density}) for the local particle density and (\ref{local-current}) for the local particle current in the following way:
\begin{eqnarray}
n(\mathbf{r}) & = & \frac{2}{\beta} \sum_{n} e^{i \omega_{n} \eta} \! \int \!\!\! \frac{d \mathbf{k}}{(2 \pi)^{3}} \,\, \mathcal{G}_{11}^{\mathrm{pf}}(\mathbf{k},\omega_{n};\mathbf{q}|\mathbf{r})
\label{local-density-pairing-fluctuations} \\
\mathbf{j}(\mathbf{r}) & = & \frac{1}{m} \left( \mathbf{Q_{0}} + \nabla \phi(\mathbf{r}) \right) n(\mathbf{r}) 
\nonumber \\
& + & \frac{2}{\beta} \sum_{n} e^{i \omega_{n} \eta} \! \int \!\!\! \frac{d \mathbf{k}}{(2 \pi)^{3}} \frac{\mathbf{k}}{m} \, \mathcal{G}_{11}^{\mathrm{pf}}(\mathbf{k},\omega_{n};\mathbf{q}|\mathbf{r}) \, .
\label{local-current-pairing-fluctuations}
\end{eqnarray}

As already mentioned, there still remains to specify the way the spatial profiles of the magnitude $|\Delta(\mathbf{r})|$ and phase $\phi(\mathbf{r})$ of the gap parameter are obtained, on whose \emph{a priori\/} knowledge the above prescriptions have been built.
This will be done in Sec.~\ref{sec:numerical_procedures}, where the expressions (\ref{local-density-pairing-fluctuations}) for $n(\mathbf{r})$ and (\ref{local-current-pairing-fluctuations}) for $\mathbf{j}(\mathbf{r})$ will serve to set up the numerical procedure for the inclusion of pairing fluctuations within the mLPDA approach.
Accordingly, in Sec.~\ref{sec:numerical_procedures} the local gap profile $\Delta(\mathbf{r})$ will be obtained from the (self-consistent) solution of the mLPDA equation, for which the condition 
$|\mathbf{j}(\mathbf{r})| = J$ for the spatial uniformity of the local particle current $\mathbf{j}(\mathbf{r})$ given by Eq.~(\ref{local-current-pairing-fluctuations})
will act as a suitable constraint on the mLPDA equation itself.

Besides being utilized in conjunction with the mLPDA equation, the expression (\ref{local-current-pairing-fluctuations}) for the local particle current is relevant in itself because it generalizes 
to the whole BCS-BEC crossover the \emph{two-fluid model at finite temperature\/}, even in the presence of spatial inhomogeneities.
In particular, Eq.~(\ref{local-current-pairing-fluctuations}) recovers Eq.~(\ref{local-current-mean-field}) in the weak-coupling (BCS) limit of strongly overlapping Cooper pairs, in line with the Bardeen fermionic
two-fluid model \cite{Bardeen-1958}.
While in the opposite strong-coupling (BEC) limit of dilute dimers, Eq.~(\ref{local-current-pairing-fluctuations}) reduces to the bosonic two-fluid model, in line with that originally introduced by Tisza \cite{Tisza-1938} and Landau \cite{Landau-1941}. 
This is explicitly shown analytically in Appendix~\ref{sec:Appendix-B} in the case of a homogeneous system.

\section{Numerical procedures} 
\label{sec:numerical_procedures}

In this Section, we describe the numerical procedures utilized for solving the differential  mLPDA equation with the inclusion of  pairing fluctuations beyond mean field.

We will specifically be concerned with  the Josephson effect, whereby a stationary supercurrent flows (say, along $x$)  across a fixed barrier that acts like  an external potential $V_{\mathrm{ext}}(x)$.
Following Ref.~\cite{PSS-2020}, here we shall consider an SsS slab geometry, with the superfluid flow along  the $x$ direction and translational invariance in  
the $y$ and $z$ directions, and with the fermionic inter-particle  attraction extending unmodified across the barrier region.
Accordingly, the value of the thermodynamic chemical potential $\mu$ and the asymptotic (bulk) value of the gap parameter $\Delta_{0}$ away from the barrier will not be affected by the presence of the
 barrier owing to its limited spatial extent.
For definiteness, the potential barrier is taken to be of the Gaussian form $V_{\mathrm{ext}}(x) = V_{0} e^{- \frac{x^{2}}{2 \sigma_{L}^{2}}}$, which is 
centered at $x = 0$ and has  height $V_{0}$ and variance $\sigma_{L}$.
By complying with typical values of condensed-matter samples \cite{footnote-Tafuri}, throughout we take $V_{0}/E_{F}=0.1$ and $k_{F} \sigma_{L}=2.5$, where $E_{F}=k_{F}^{2}/(2m)$ is the Fermi energy
and $k_{F}=(3 \pi^{2} n)^{1/3}$ the Fermi wave vector with density $n$.

When dealing with a Fermi gas with attractive inter-particle interaction like in the present case, a common practice is to span the corresponding BCS-BEC crossover in terms of the dimensionless coupling $(k_{F} a_{F})^{-1}$, where $a_{F}$ is the scattering length of the two-fermion problem \emph{in vacuum\/} \cite{Physics-Reports-2018}.
This  coupling parameter ranges from $(k_{F}\, a_{F})^{-1} \lesssim -1$ in the weak-coupling (BCS) regime when $a_{F} < 0$, to $(k_{F}\, a_{F})^{-1} \gtrsim +1$ in the strong-coupling (BEC) regime when $a_{F} > 0$, across the unitary limit $(k_{F}\, a_{F})^{-1} = 0$ when $|a_{F}|$ diverges.

While Ref.~\cite{PSS-2020} was mostly focused on  the BCS side of unitarity, with the argument  that pairing fluctuations were not included in that reference  like in the original LPDA approach of Ref.~\cite{SS-2014}, here we will instead be able to  explore the whole  BCS-BEC crossover, once pairing fluctuations will be properly   included on top of the LPDA approach.

\begin{center}
{\bf A. Solving for the differential LPDA \\ and mLPDA equations}
\end{center}

With these premises, the differential LPDA equation introduced in Ref.~\cite{SS-2014} takes the form  considered in Ref.~\cite{PSS-2020} for the Josephson effect
with an effectively one-dimensional geometry:
\begin{eqnarray}
-\dfrac{m}{4\pi a_F}\tilde{\Delta}(x) & = &\mathcal{I}_0(x)\tilde{\Delta}(x)+\dfrac{\mathcal{I}_1(x)}{4m}\dfrac{d^2}{dx^2}\tilde{\Delta}(x)
\nonumber \\
& + & i  \, \mathcal{I}_1(x) \dfrac{Q_0}{m} \frac{d \tilde{\Delta}(x)}{dx} \, .
\label{LPDA-equation}
\end{eqnarray}
Here, $\tilde{\Delta}(x)=e^{-2iQ_{0}x}\Delta(x) \equiv |\tilde{\Delta}(x)| \, e^{2 i \phi(x)}$ with $Q_{0} = |\mathbf{Q}_{0}|$ (cf. Eq.~(\ref{replacement-phase-Delta})),
and the coefficients $\mathcal{I}_0(x)$ and $\mathcal{I}_1(x)$ are given by the expressions \cite{SS-2014} 
\begin{eqnarray}
\mathcal{I}_{0}(x) & = & \int \! \frac{d \mathbf{k}}{(2 \pi)^{3}} \, 
\left\{ \frac{ 1 - 2 f_{F}(E_{+}^{\mathbf{Q}_{0}}(\mathbf{k}|x)) }{2 \, E(\mathbf{k}|x)} - \frac{m}{\mathbf{k}^{2}} \right\}
\label{I_0} \\
\mathcal{I}_{1}(x) & = & \frac{1}{2} \, \int \! \frac{d \mathbf{k}}{(2 \pi)^{3}} 
\left\{  \frac{\xi(\mathbf{k}|x)}{2 \, E(\mathbf{k}|x)^{3}} \, \left[  1 - 2 f_{F}(E_{+}^{\mathbf{Q}_{0}}(\mathbf{k}|x)) \right] \right.                          
\nonumber \\
& + &  \frac{\xi(\mathbf{k}|x)}{E(\mathbf{k}|x)^{2}} \, 
\frac{\partial f_{F}(E_{+}^{\mathbf{Q}_{0}}(\mathbf{k}|x))}{\partial E_{+}^{\mathbf{Q}_{0}}(\mathbf{k}|x)} 
\nonumber \\
& + & \left. \frac{\mathbf{k}\cdot\mathbf{Q}_{0}}{\mathbf{Q}_{0}^{2}} \, \frac{1}{E(\mathbf{k}|x)} \, 
\frac{\partial f_{F}(E_{+}^{\mathbf{Q}_{0}}(\mathbf{k}|x))}{\partial E_{+}^{\mathbf{Q}_{0}}(\mathbf{k}|x)} \right\} 
\label{I_1}
\end{eqnarray}
where
\begin{eqnarray}
\xi(\mathbf{k}|x) & = & \frac{\mathbf{k}^{2}}{2m} - \mu + V_{\mathrm{ext}}(x) +  \frac{\mathbf{Q}_{0}^{2}}{2m}
\label{notation-LPDA-I} \\
E(\mathbf{k}|x) & = & \sqrt{ \xi(\mathbf{k}|x)^{2} + |\tilde{\Delta}(x)|^{2} }
\label{notation-LPDA-II} \\
E_{+}^{\mathbf{Q}_{0}}(\mathbf{k}|x) & = & E(\mathbf{k}|x) + \frac{\mathbf{k} \cdot \mathbf{Q}_{0}}{m} \, .
\label{notation-LPDA-III} 
\end{eqnarray}
Note that the expressions (\ref{notation-LPDA-I})-(\ref{notation-LPDA-III}) correspond to the expressions (\ref{notation-I-inhomogeneous})-(\ref{notation-III-inhomogeneous}) where one sets $ \nabla \phi(\mathbf{r}) = 0$,
in accordance with the original derivation of the LPDA equation given in  Ref.~\cite{SS-2014}.
Note further that the expression (\ref{notation-LPDA-I}) is the only place where the external potential $V_{\mathrm{ext}}(x)$ explicitly appears in the context of the mLPDA approach 
(while in the context of the LPDA approach $V_{\mathrm{ext}}(x)$ enters also  the expressions (\ref{local-density-mean-field}) for the local density and (\ref{local-current-mean-field}) for the local current 
at the mean-field level).
Note, finally, that full three-dimensional geometry in which the physical system is embedded (which is essential to account for the three-dimensional structure of Cooper pairs and/or dimers)
explicitly appears in the wave-vector integrals of Eqs.~(\ref{I_0}) for $\mathcal{I}_{0}$ and (\ref{I_1}) for $\mathcal{I}_{1}$.

Following Appendix A of Ref.~\cite{PSS-2020}, the LPDA \emph{second order\/} differential equation (\ref{LPDA-equation}) can be  
reduced to a system of four \emph{first-order\/} differential equations, by separating its real and imaginary parts associated with the complex function $\tilde{\Delta}(x) = |\tilde{\Delta}(x)| \, e^{2 i \phi(x)}$ in the presence of a supercurrent  and introducing the spatial derivatives of $|\tilde{\Delta}(x)|$ and $\phi(x)$.
In addition, the imaginary part of the LPDA equation is conveniently  replaced by the requirement 

\begin{equation}
j(x) = J
\label{crucial-requirement}
\end{equation}

for the local particle current to be everywhere uniform.
This replacement is common practice when solving for the GL equation close to the critical temperature \cite{Jacobson-1965} or for the GP equation at zero temperature \cite{Hakim-1997}.
More generally, even when solving numerically up to self-consistency the BdG equations throughout the BCS-BEC crossover at zero temperature,
one finds it convenient to replace the imaginary part of the gap equation therein with the requirement for the local current to be everywhere uniform \cite{SPS-2010}.

In Ref.~\cite{PSS-2020} the requirement (\ref{crucial-requirement})  was enforced by taking the local particle current $j(x)$ of the mean-field form (\ref{local-current-mean-field}),
in line with the original  LPDA approach of Ref.~\cite{SS-2014}.
Here, we improve on that approach and take instead the local particle current $j(x)$ of the form (\ref{local-current-pairing-fluctuations}), which includes beyond-mean-field pairing fluctuations in the presence of a supercurrent in a consistent way.

This replacement of the expression of the local particle current (from Eq.~(\ref{local-current-mean-field}) to Eq.~(\ref{local-current-pairing-fluctuations})) is the key step for obtaining what is referred to as 
the \emph{modified\/} LPDA (or mLPDA) approach, which includes pairing fluctuations beyond mean field and represents the main contribution of the present article.
With this replacement, one can solve for the mLPDA differential equation by adopting  otherwise the same numerical procedures described in Appendix A of Ref.~\cite{PSS-2020}.
One should also recall that, within the mLPDA approach, the expressions (\ref{I_0}) for $\mathcal{I}_{0}(x)$ and (\ref{I_1}) for $\mathcal{I}_{1}(x)$ contain the thermodynamic chemical potential 
$\mu$ which now consistently  includes the effect of pairing fluctuations via the density equation (\ref{local-density-pairing-fluctuations}).

\begin{center}
{\bf B. Solving for the particle-particle ladder \\  in the presence of a supercurrent}
\end{center}

We now describe the numerical procedure we have followed for calculating the expressions (\ref{full-particle-particle-ladder})-(\ref{B}) for the particle-particle ladder $\Gamma_{11}(Q;\mathbf{q})$ and (\ref{self-energies-t-matrix}) of the diagonal self-energy $\mathfrak{S}_{11}^{\mathrm{pf}}(k;\mathbf{q})$ within the $t$-matrix approach in the presence of a supercurrent.
These quantities  are required to obtain the local particle current (\ref{local-current-pairing-fluctuations}) with the inclusion of pairing fluctuations needed 
for solving the mLPDA equation.
To this end, we adapt to the present context  the procedures utilized in Ref.~\cite{PPS-2004}, where these calculations were originally performed 
in the homogeneous case only and in the absence of a supercurrent.

To begin with, special care has to be used when handling the sums over the bosonic Matsubara frequencies $\Omega_{\nu}$ in Eq.~(\ref{self-energies-t-matrix}) and over the fermionic Matsubara frequencies $\omega_{n}$ in Eqs.~(\ref{local-density-pairing-fluctuations}) and (\ref{local-current-pairing-fluctuations}) (the sum over the fermionic Matsubara frequencies in Eqs.~(\ref{A})-(\ref{B}), on the other hand, 
can be done analytically due to the presence therein of the mean-field expressions (\ref{simplified-solutions-mean-field-11}) and (\ref{simplified-solutions-mean-field-12}) for the single-particle Green's functions).
As it was done in Ref.~\cite{PPS-2004}, the sum over the bosonic Matsubara frequencies in Eq.~(\ref{self-energies-t-matrix}) is conveniently handled by introducing a cutoff $\Omega_{c}$, 
past which the sum is transformed into an integral over a real frequency $\Omega$ using the asymptotic form of $\Gamma_{11}(Q;\mathbf{q})$ (cf., e.g., Eq.~(43) of Ref.~\cite{PPS-2004} with $z \rightarrow \Omega$).
To speed up instead the sum over the fermionic Matsubara frequencies in Eqs.~(\ref{local-density-pairing-fluctuations}) and (\ref{local-current-pairing-fluctuations}), it is sufficient to add and subtract the corresponding expressions for the particle density and current at the mean-field level.

In addition, while the angular integration over the bosonic wave vector $\mathbf{Q}$ in expression (\ref{self-energies-t-matrix}) of the diagonal self-energy $\mathfrak{S}_{11}^{\mathrm{pf}}(k;\mathbf{q})$ 
 can be dealt with analytically, by making the polar angle  
to appear only inside the expression of $G_{11}^{\mathrm{mf}}$ therein by a suitable choice of the integration axes, 
in the expressions (\ref{diagonal-bubble})-(\ref{off-diagonal-bubble}) of the particle-particle bubbles the additional angular integration over the fermionic wave vector $\mathbf{k}$, that arises  
from the presence of the wave vector $\mathbf{q}$ associated with the supercurrent, has instead to be dealt with numerically.

Finally, the dependence on the spatial coordinate $\mathbf{r}$, which is required to obtain the local particle current $\mathbf{j}(\mathbf{r})$ needed for the numerical solution of the mLPDA equation, is introduced in the
expression of $\mathfrak{S}_{11}^{\mathrm{pf}}(k;\mathbf{q})$ via the prescriptions described in Sec.~\ref{sec:theoretical_approach}-E.

All these calculations are repeated at each cycle of self-consistency during the numerical solution of the mLPDA equation, with the magnitude and phase of the gap parameter 
being consistently updated at each cycle. 

\begin{figure}[t]
\begin{center}
\includegraphics[width=6.5cm,angle=0]{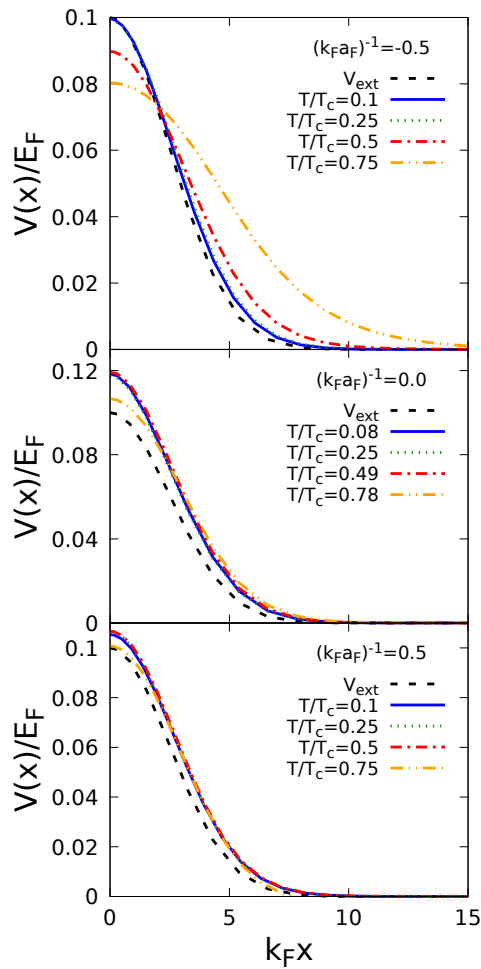}
\caption{Effective potential $V_{\mathrm{eff}}(x)$ (in units of the Fermi energy $E_{F}$) associated with an external Gaussian potential $V_{\mathrm{ext}}(x) = V_{0} e^{- \frac{x^{2}}{2 \sigma_{L}^{2}}}$ 
              with $V_{0}/E_{F}=0.1$ and $k_{F} \sigma_{L}=2.5$ (broken line).
              The spatial profiles of $V_{\mathrm{eff}}(x)$ are shown for three couplings across the BCS-BEC crossover and several temperatures, with  $Q_{0}$ taken at the maximum 
              of the Josephson characteristics in each case.}
\label{Figure-2}
\end{center} 
\end{figure}  

\begin{center}
{\bf C. Connecting the two above procedures}
\end{center}

To speed up the interconnection between the two numerical codes, one for the solution of the mLPDA equation (cf. Sec.~\ref{sec:numerical_procedures}-A) 
and the other one for the calculation of the diagonal self-energy $\mathfrak{S}_{11}^{\mathrm{pf}}$ within the $t$-matrix approximation (cf. Sec.~\ref{sec:numerical_procedures}-B), 
it is convenient to  somewhat modify the numerical procedures previously utilized in Ref.~\cite{PSS-2020} (cf. Sec. 4 of the Appendix therein) for decreasing  the numerical noise when solving for  the LPDA equation.  

Specifically, when solving for  the LPDA equation in Ref.~\cite{PSS-2020}, the expressions of ${\mathcal{I}_{0}}$ and ${\mathcal{I}_{1}}$ as well as those of the local number density and current were accurately evaluated over a regular grid of $100 \times 100 \times 100$ points in the variables ($|\Delta|,\mu,Q_{0}+\frac{d \phi(x)}{dx}$), and a linear interpolation was then utilized in each variable to determine the actual values of the quantities needed at each iteration of the Newton method.
Here, when solving for  the mLPDA equation, this procedure becomes impractical due  to the time required when evaluating the expressions of the density and current with the inclusion of pairing fluctuations.

To work around this problem, we have kept the above grid of $100 \times 100 \times 100$ points in the variables ($|\Delta|,\mu,Q_{0}+\frac{d \phi(x)}{dx}$) supplemented as before by a linear interpolation for calculating ${\mathcal{I}_{1}}$  and ${\mathcal{I}_{0}}$;
but at the same time  we have drastically reduced the size of the grid for calculating the density and current to $5 \times 5$ points only in the variables ($|\Delta|,Q_{0}+\frac{d \phi(x)}{dx}$), now supplemented by a more sophisticated 2D spline interpolation in each variable.
Eliminating here the chemical potential $\mu$ from the grid is connected with the procedure described in Sec.~\ref{sec:theoretical_approach}-E, which takes care of the local chemical potential in the process of calculating the local effective potential.
In addition, back to the context of the LPDA equation, we have verified that a grid of $5 \times 5 \times 5$ points in the variables ($|\Delta|,\mu,Q_{0}+\frac{d \phi(x)}{dx}$) supplemented by a 3D spline interpolation in each variable would suffice  to obtain the density and current with good enough  accuracy, when compared with the more lengthy calculations with $100 \times 100 \times 100$ points made originally in Ref.~\cite{PSS-2020}.

\section{Numerical results} 
\label{sec:numerical_results}

In this Section, we report on  a number of numerical results obtained by solving the LPDA and mLPDA equations on equal  footing.
Specifically, this will be done by considering a  simple Josephson configuration, where a potential barrier $V_{\mathrm{ext}}(x)$ with a slab geometry of the SsS type is embedded in a otherwise  homogeneous fermionic superfluid which extends  to infinity on both sides of the barrier.
Detailed consideration of more complex Josephson spatial configurations, like those corresponding to the experiments of Refs.~\cite{Kwon-2020,DelPace-2021}, is instead given in the companion article \cite{PPS-companion-PRL-article}.

\begin{figure}[t]
\begin{center}
\includegraphics[width=7.5cm,angle=0]{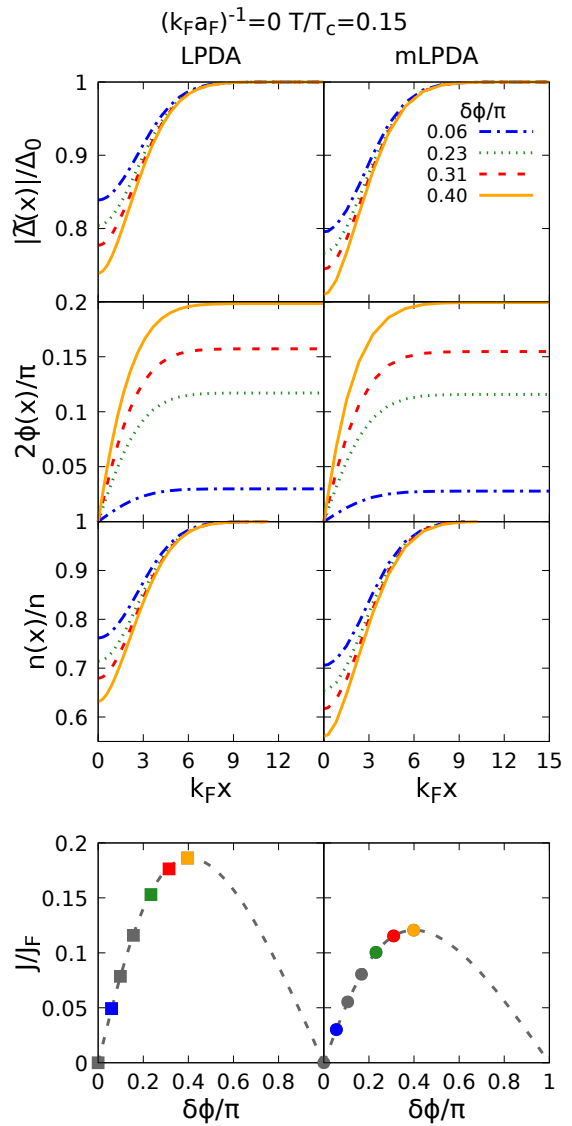}
\caption{From top to bottom: Spatial profiles of the  magnitude $\tilde{|\Delta}(x)|$ and phase $2 \phi(x)$ of the gap parameter and of the  density $n(x)$, 
             and Josephson characteristics $J(\delta\phi)$  vs  the asymptotic phase difference $\delta \phi$,  as  obtained by solving the LPDA equation 
             (left panels) and the mLPDA equation (right panels) at unitarity and $T/T_{c}=0.15$, with the same barrier of Fig.~\ref{Figure-2}.
             The bulk values $\Delta_{0}$ of the (magnitude of the) gap parameter and $n$ of the density are used for normalization, together with $J_{F}=k_{F}n/m$ for the current.
             Here and in the following figures, the temperature $T$ is normalized to the respective value of $T_{c}$ obtained within either  the  LPDA or mLPDA approaches.}
\label{Figure-3}
\end{center} 
\end{figure}  

\begin{center}
{\bf A. Effective vs external potentials \\ when including local pairing fluctuations}
\end{center}

In Sec.~\ref{sec:theoretical_approach}-E we have argued that, when solving for the mLPDA equation, one needs to introduce an effective potential $V_{\mathrm{eff}} (\mathbf{r})$ in the diagonal element 
$\Gamma_{11}(Q;\mathbf{q}|\mathbf{r})$ of the particle-particle ladder, in the place of the external potential $V_{\mathrm{ext}} (\mathbf{r})$ corresponding to  the geometrical constraints in which the Fermi
 superfluid is embedded.
We have also concluded that, by construction, the effective potential $V_{\mathrm{eff}} (\mathbf{r})$ depends not only on the external potential
$V_{\mathrm{ext}} (\mathbf{r})$, but also  on coupling, temperature, and the wave vector $\mathbf{Q}_{0}$ associated with the supercurrent.

Figure~\ref{Figure-2}  shows a number of profiles of $V_{\mathrm{eff}}(x)$  associated with  a given $V_{\mathrm{ext}}(x)$ with the one-dimensional geometry 
considered throughout, for three couplings across the BCS-BEC crossover and several temperatures in the superfluid phase.
[It is sufficient to show $V_{\mathrm{eff}}(x)$ only for $x>0$ due to reflection  symmetry.]
In all  cases, the value of $Q_{0}$ at which $V_{\mathrm{eff}}(x)$ is calculated (cf. Sec.~\ref{sec:theoretical_approach}-E) corresponds to the maximum of the associated Josephson characteristic
(cf. the bottom of  Fig.~\ref{Figure-3}  below).
Note that the differences between $V_{\mathrm{eff}}(x)$ and $V_{\mathrm{ext}}(x)$ are never  too pronounced, and that they become most noticeable at the highest temperature.
Note also the monotonic behavior for increasing temperature at each coupling.

\begin{figure}[t]
\begin{center}
\includegraphics[width=8.2cm,angle=0]{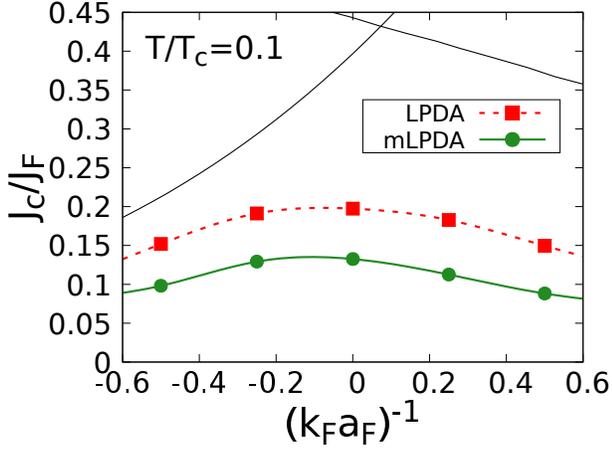}
\caption{Coupling dependence of the critical current $J_{c}$ (in units of $J_{F}$) obtained in the crossover region for $T/T_{c}=0.1$  within the LPDA and mLPDA approaches, 
              with the same barrier of Fig.~\ref{Figure-2}.
              The upper  left and right full  curves correspond to the appearance, respectively, of pair-breaking and sound-mode excitations at zero temperature in the limit of a vanishing barrier (like in Fig.~24 of Ref.~\cite{SPS-2010}).}
\label{Figure-4}
\end{center} 
\end{figure}  

\begin{center}
{\bf B. Comparison between the solutions of \\ the LPDA and mLPDA equations}
\end{center}

A direct comparison between the solutions of the LPDA and mLPDA equations is shown in Fig.~\ref{Figure-3}, for the same physical barrier considered in 
 Fig.~\ref{Figure-2}. 
Here, the spatial profiles of the magnitude $|\tilde{\Delta}(x)|$ and phase $\phi(x)$ of the gap parameter contributed by  the presence of the barrier are shown together with the local density $n(x)$ at unitarity for the temperature $T/T_{c}=0.15$, for the four (out of six) values of the asymptotic phase difference $\delta \phi$ accumulated across the barrier which are reported with the same colors in the Josephson characteristics of the bottom panels.
Note how the inclusion of pairing fluctuations implemented by the mLPDA equation  acts to decrease  the local values of the magnitude of the gap parameter and of the density inside 
 as well as  close to the barrier, with respect to those obtained by the LPDA equation.
This  decrease  is also evident for the critical value $J_{c}$ of the current, that corresponds to the maximum of the Josephson characteristics in the bottom panels of Fig.~\ref{Figure-3}.

\begin{center}
{\bf C. Further comparison of the critical currents \\ vs coupling and temperature}
\end{center}

The above decrease  of $J_{c}$  when  including pairing fluctuations persists at low enough temperature  over the whole  crossover region 
on both (BCS and BEC) sides of unitarity.
This is shown in Fig.~\ref{Figure-4}  with the same barrier  considered  in the previous figures and for $T/T_{c}=0.1$, 
 where the values of $T_{c}$ are calculated for each coupling within the LPDA and mLPDA approaches, respectively.
For this barrier, the maximum value of $J_{c}$ is attained at (about) unitarity in both (LPDA and mLPDA) cases, thus confirming the result obtained in Ref.~\cite{SPS-2010} 
by a self-consistent solution of the BdG equations at zero temperature with a comparable barrier (cf. Fig.~25 therein).

\begin{figure}[t]
\begin{center}
\includegraphics[width=7.2cm,angle=0]{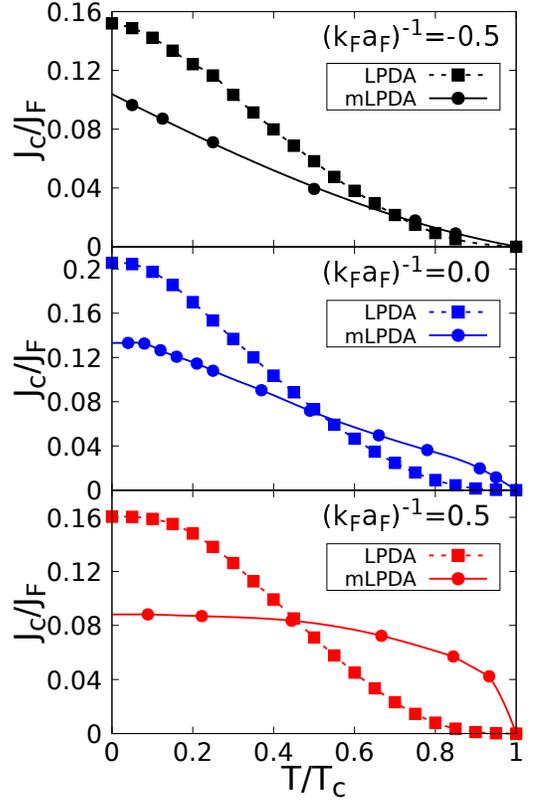}
\caption{Temperature dependence of the critical current $J_{c}$ (in units of $J_{F}$), obtained within the LPDA (filled squares)  and mLPDA (filled circles)  approaches 
              for three couplings spanning the BCS-BEC crossover, with the same barrier of Fig.~\ref{Figure-2}.}
\label{Figure-5}
\end{center} 
\end{figure}  

The situation can, however, get reversed for increasing temperature. 
This is shown in Fig.~\ref{Figure-5}  where the temperature dependence of $J_{c}$, obtained by solving the LPDA and mLPDA equations from $T=0$ up to $T=T_{c}$, is reported for three couplings 
across the BCS-BEC crossover. 
(Also in this case, $T_{c}$ is calculated for each coupling within the LPDA and mLPDA approaches, respectively.) 
Note that a crossing between the LPDA and mLPDA results occurs at an intermediate  temperature, whose value relative to $T_{c}$  decreases as the coupling evolves
from the BCS to the BEC regimes.
Note also that, within the mLPDA approach, the temperature dependence of $J_{c}$  changes from a convex to a concave behavior from the BCS to the BEC regime, passing through an essentially linear behavior at unitarity.
This result will be relevant  when interpreting the results shown in Fig.~4 of the companion article \cite{PPS-companion-PRL-article},  which are  obtained still at unitarity
but with a much more complex spatial geometry corresponding to experiments with ultra-cold Fermi gases.

\vspace{1.5cm}

\begin{center}
{\bf D. A preliminary inclusion of \\ the extended GMB contribution}
\end{center}

Throughout the present article, we have implemented the inclusion of pairing fluctuations over and above the LPDA approach of Ref.~\cite{SS-2014}, by resting on the non-self-consistent $t$-matrix approach 
in the superfluid phase of Ref.~\cite{PPS-2004}.
Here, the approach of Ref.~\cite{PPS-2004} has suitably been adapted to describe  spatially inhomogeneous situations  as well as to  include 
the presence of a supercurrent.

\begin{figure}[t]
\begin{center}
\includegraphics[width=8.8cm,angle=0]{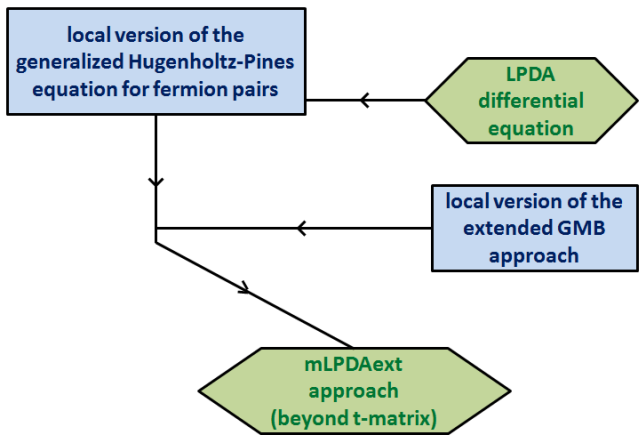}
\caption{``Flow diagram'' showing schematically the way the mLPDAext approach is set up. 
              At a first step, the LPDA equation is cast in the framework of the generalized Hugenholtz-Pines gap equation of Ref.~\cite{PPSb-2018}, now within a local perspective.
              At a second step, this equation is merged with a local version of the extended GMB approach of Ref.~\cite{PPSb-2018}, yielding eventually the mLPDAext approach.}
\label{Figure-6}
\end{center} 
\end{figure}  

In the homogeneous case, on the other hand, a quite general procedure was proposed in Ref.~\cite{PPSb-2018} for  a systematic inclusion of beyond-mean-field pairing fluctuations in the gap equation.
In particular, in Ref.~\cite{PPSb-2018} this procedure was  implemented by extending to the whole BCS-BEC crossover the Gorkov-Melik-Barkhudarov (GMB) contribution in the broken-symmetry phase \cite{GMB-1961}, which was originally considered in the BCS limit only.
A Popov-like correction \cite{PS-2005} was further included in Ref.~\cite{PPSb-2018}, to account for some degree of self-consistency in the $t$-matrix approach of Ref.~\cite{PPS-2004}.
Taken together, the GMB and Popov contributions, once properly extended to the whole BCS-BEC crossover, were dubbed the ``extended GMB approach''.
The numerical results obtained by this approach have recently been validated throughout the BCS-BEC crossover  in different experimental contexts, 
namely,  at low temperature in the superfluid phase \cite{Moritz-2022} as well as at the critical temperature \cite{Koehl-2023}.

In principle, no impediment should appear in further adapting this extended GMB approach to spatially inhomogeneous situations, aiming at  further improving on the mLPDA approach
 as anticipated in Sec.~\ref{sec:introduction}.
Figure~\ref{Figure-6}   schematically summarizes the way how a \emph{local\/} version of the extended GMB approach could be combined with the differential LPDA equation, 
resulting in an ``extended'' version of the mLPDA approach which may shortly be referred to as mLPDAext approach.
In practice, however, implementing this mLPDAext approach is expected to require substantial theoretical and computational efforts (especially in the presence of a supercurrent), which by far exceed 
the aims of the present work. 
Nonetheless, we can here provide a preliminary yet meaningful demonstration of how the inclusion of the extended GMB correction may influence a physical quantity 
like the superfluid density $\rho_{s}$, of interest to  the present context. 

\begin{figure}[t]
\begin{center}
\includegraphics[width=8.85cm,angle=0]{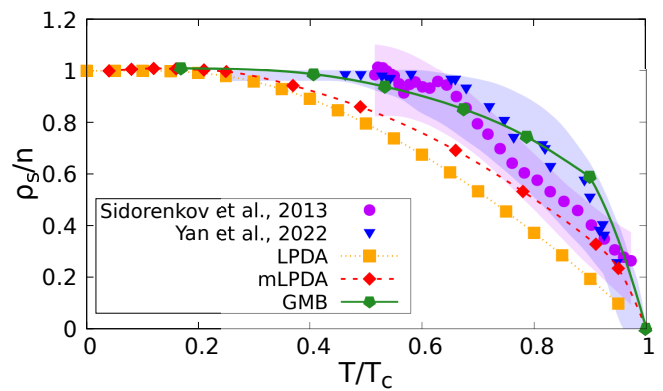}
\caption{Temperature dependence of the superfluid density at unitarity.
              The results of the LPDA (squares) and mLPDA (diamonds) approaches  are compared with the experimental data from Ref.~\cite{Sidorenkov-2013} (dots) and Ref.~\cite{Yan-2022} (triangles),
              where the shaded regions stand for the corresponding experimental uncertainties.
              Preliminary results obtained with the extended GMB approach of Ref.~\cite{PPSb-2018} are also shown (pentagons).}
\label{Figure-7}
\end{center} 
\end{figure}  
 
 To this end, we calculate the current $j$ without the presence of a  barrier for an infinitesimal (in practice, quite small) value of $Q_{0}$, such that $\rho_{s}$  is  
 obtained from the relation $j = \rho_{s} Q_{0}/m$ for various couplings and temperatures.
 We do this, first at the level of the LPDA approach using for $j$ the expression (\ref{local-current-mean-field}), and then at the level of the mLPDA approach using for $j$ the expression (\ref{local-current-pairing-fluctuations}).
The corresponding temperature dependences of $\rho_{s}$ at unitarity are shown in Fig.~\ref{Figure-7}, where the experimental results from Ref.~\cite{Sidorenkov-2013} and 
Ref.~\cite{Yan-2022} are also reported for comparison.
Although the mLPDA approach is seen to improve  this comparison with respect to the LPDA approach, there still appear noticeable discrepancies with the experimental results  
which call for further improvements on the mLPDA approach.

This improvement can be achieved in a preliminary fashion, by adapting to the present context the  simplified procedure described in Ref.~\cite{PPPS-2018} (as summarized in the paragraph following Eq.~(47) therein), whereby passing from the non-self-consistent $t$-matrix approach to the GMB approach amounts  essentially  to introducing  a suitable temperature-dependent shift on  the value of the coupling, which can be calculated in a well defined way.
The effect of this procedure on the temperature dependence of the superfluid density at unitarity is also reported in Fig.~\ref{Figure-7}, resulting in  a definite improvement 
on the comparison with the experimental data.

A similar procedure is also applied in Fig.~3(a) of Ref.~\cite{PPS-companion-PRL-article} to four data points obtained by the mLPDA approach on the BCS side of unitarity.
In this case, the procedure is  applied to each of the tubular filaments in which the (nearly) cylindrical trap is  partitioned, with the favorable outcome that essentially a common shift for the value of the coupling proves sufficient (to within a $10 \%$ uncertainty).
Even in this case, this simplified version of the extended GMB approach has produced a non-negligible improvement over the mLPDA approach  when comparing with the experimental data.

\section{Concluding remarks and perspectives}
\label{sec:conclusions}

In this article, we have set up  and implemented  the mLPDA  approach that allows for the inclusion of pairing fluctuations beyond mean field in inhomogeneous  superfluid Fermi systems, with the purpose of dealing  with problems of physical and experimental interest for which  the effects of non-trivial geometrical constraints are as important as those  originating from  the inter-particle dynamics and have thus  to be treated on  equal  footing.
The BCS-BEC crossover occurring in ultra-cold trapped  Fermi gases is an ideal setting for testing this type of approach, since it involves spanning a wide region of coupling and temperature that can be reasonably dealt with only by a sufficiently broad-ranging theoretical approach.

To this end, the strategy we have adopted has rested on  including  pairing fluctuations directly on the  differential  LPDA equation of Ref.~\cite{SS-2014},
 by combining this equation  with elements drawn from Ref.~\cite{PPS-2004}.
This procedure resulted in what we have referred to as  the mLPDA  approach.
In particular, this approach was set up in the presence of a supercurrent, which makes the theoretical setup unavoidably more involved but permits applications to problems  of wide interest like those involving  the Josephson effect.
In the present article these problems have specifically been considered for a geometry that mostly applies to condensed-matter samples, while  in the companion article \cite{PPS-companion-PRL-article} a more complex geometry appropriate to ultra-cold trapped Fermi gases is considered.

The above strategy  may acquire an even broader importance, when  connecting it with the approach developed in Ref.~\cite{PPSb-2018}.
There  the gap equation in the homogeneous case was cast in the form of a Hugenholtz-Pines condition for fermion pairs, which  allows for a systematic inclusion of pairing-fluctuation corrections beyond mean field over and above the $t$-matrix approximation considered in the present article.
This is because in Ref.~\cite{PPSb-2018} the gap equation was related  to the two-particle Green's function in the superfluid phase, instead of following the common practice of
associating it with the single-particle Green's function \cite{Nozieres-1964}.
Taking then advantage of the fact that  the LPDA approach extends \emph{directly\/} the equation for the gap parameter to the inhomogeneous case, the  approach  
of Ref.~\cite{PPSb-2018} could  be formally taken over to the inhomogeneous case  so as to include in the LPDA approach pairing-fluctuation corrections 
even  \emph{beyond\/} the $t$-matrix approximation.
In this way, one may go even beyond the mLPDA approach,  yielding what in Sec.\ref{sec:numerical_results}-D is referred to as the extended mLPDAext approach.
Favorable preliminary results along these lines have been  presented in the present article as well as the companion article \cite{PPS-companion-PRL-article}, 
by relying on  the extended GMB approach of Ref.~\cite{PPSb-2018}.
However, extensive  implementation of this beyond mLPDA strategy appears highly non-trivial (especially in the presence of a supercurrent) and requires further dedicated projects to which it is postponed.

It should be emphasized that, irrespective of whether one stops at the level of the mLPDA approach (as we have done in the present article) or plans for further improvements (like the extended mLPDAext approach 
outlined above), our whole strategy for including pairing fluctuations in the context of the LPDA differential equation of Ref.~\cite{SS-2014} is, by construction, ``modular'' in nature since it rests on a
progressive inclusion of diagrammatic terms borrowed from the many-body Green's functions theory.
Accordingly, these diagrammatic terms do not contain ``internal'' parameters whose values need to be fitted by utilizing the results of independent theoretical calculations, as it is instead the case with approaches based on Density Functional Theory \cite{Scuseria-2005}.
In this respect, numerical results obtained by the mLPDA approach should be considered as \emph{first-principle\/} calculations.

However, when comparing available experimental data with theoretical results, obtained even in the absence of internal parameters like in our case, the unavoidable occurrence of experimental uncertainties introduces
a source of ``external'' parameters in the theoretical calculations, which have accordingly to be taken into account for the sake of the comparison.
For instance, in experiments with ultra-cold trapped gases the uncertainty in the total number of atoms $N$ can be as large as $30 \%$, such that theoretical results aiming at comparing with experimental data for
these systems should consider $N$ like an external parameter, as it is done in the companion article \cite{PPS-companion-PRL-article}.
Note that analogous problems would occur even in condensed matter when performing successful first-principle band-structure calculations based on the many-body Green's functions theory \cite{Strinati-1980}, 
if, for instance, the lattice constant would not be known experimentally with sufficient accuracy (which, in practice, is not the case).


\begin{center}
\begin{small}
{\bf ACKNOWLEDGMENTS}
\end{small}
\end{center}

We are indebted to G. Roati for support and discussions.
Partial financial support from the Italian MIUR under Project PRIN2017 (20172H2SC4) is acknowledged.

\newpage
\appendix   
\vspace{1.5cm}
\section{PARTICLE-PARTICLE LADDER \\ FOR THE SUPERFLUID PHASE \\ IN THE PRESENCE OF A SUPERCURRENT}
\label{sec:Appendix-A}

In this Appendix, we explicitly implement the form (\ref{self-energies-t-matrix}) of the self-energy that includes pairing fluctuations in the presence of a stationary supercurrent within the $t$-matrix approach.
To this end, it is sufficient to construct the diagonal component $\Gamma_{11}$ of the particle-particle ladder in  the broke-symmetry phase, since the normal single-particle Green's function
$\mathcal{G}_{11}^{\mathrm{mf}}$ needed in Eq.~(\ref{self-energies-t-matrix}) is given by Eq.~(\ref{simplified-solutions-mean-field-11}).
In this way, we shall  also verify that the properties (\ref{G11-and-q-vector})-(\ref{G22-and-q-vector}) hold specifically for all components of the self-energy within the $t$-matrix approximation of interest.
The procedure for extending these results to the presence of spatial inhomogeneities, which  arise  from an external potential (like a barrier for the Josephson effect), 
is implemented in Sec.~\ref{sec:theoretical_approach}-E. 

We begin by briefly recalling the procedure described in Appendix A of Ref.~\cite{APS-2003}, to obtain the elements $\Gamma_{ii'}$ of the particle-particle ladder in  the broken-symmetry phase.
This procedure rests on the choice of a contact potential with (negative) strength $v_{0}$ to represent the inter-particle attraction, which entails a suitable regularization in terms of a cutoff $k_{0}$ and 
of the  scattering length $a_{F}$ of the two-fermion problem, in the form \cite{Physics-Reports-2018}:
\begin{equation}
\frac{m}{4 \pi a_{F}} = \frac{1}{v_{0}} + \int_{|\mathbf{k}| \le k_{0}} \! \frac{d \mathbf{k}}{(2 \pi)^{3}} \, \frac{m}{\mathbf{k}^{2}} \, .
\label{regularization}
\end{equation}
By this regularization, the limits $v_{0} \rightarrow 0^{-}$ and $k_{0} \rightarrow \infty$ are taken simultaneously so that $a_{F}$ is kept at the desired value.
In this way, only selected classes of diagrams of the many-body diagrammatic structure survive the regularization (\ref{regularization}) when the limit $v_{0} \rightarrow 0^{-}$ is taken.

\begin{figure}[t]
\begin{center}
\includegraphics[width=7.5cm,angle=0]{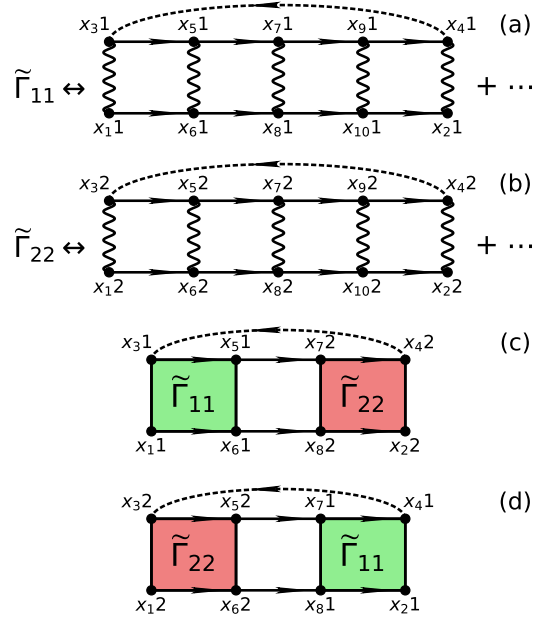}
\caption{(Color online)  Diagrams for the particle-particle ladder in $x$-space, in  the presence of a stationary supercurrent.
                                      Examples are shown for: (a) the 11-component, (b) the 22-component, (c) the 21-component, and (d) the 12-component.
                                      Full lines stand for the fermionic single-particle Green's functions $G_{ii'}(x,x')$ and wiggly lines for the contact inter-particle attraction.
                                      At each vertex (dot), coordinates ($x$) and Nambu indices ($i$) are indicated.
                                      In each diagram, an upper single-particle (dashed) line is appended which closes the fermion loop, thus obtaining  the self-energy $\Sigma_{ii'}(x_{2},x_{1})$ 
                                      within the $t$-matrix approximation.}
\label{Figure-8}
\end{center} 
\end{figure} 

Specifically, for the the particle-particle ladder of interest here it was shown in Appendix A of Ref.~\cite{APS-2003} that only ladder diagrams constructed by pairs of either $G_{11}$ or $G_{22}$ single-particle Green's functions survive the regularization (\ref{regularization}).
Examples of these diagrams in $x$-space are shown in Figs.~\ref{Figure-8}(a)  and \ref{Figure-8}(b).
In the presence of a stationary supercurrent, owing to the properties (\ref{G11-and-q-vector})-(\ref{G22-and-q-vector}) of the single-particle Green's functions, one readily realizes that 
$\tilde{\Gamma}_{11}$ of Fig.~\ref{Figure-8}(a)  acquires the phase factor $e^{i \mathbf{q} \cdot (\mathbf{r}_{2} + \mathbf{r}_{4} - \mathbf{r}_{1} - \mathbf{r}_{3})}$, while
$\tilde{\Gamma}_{22}$ of Fig.~\ref{Figure-8}(b)  acquires the phase factor $e^{- i \mathbf{q} \cdot (\mathbf{r}_{2} + \mathbf{r}_{4} - \mathbf{r}_{1} - \mathbf{r}_{3})}$.
In addition, structures like $\tilde{\Gamma}_{11}$ and $\tilde{\Gamma}_{22}$ can be mutually connected through rungs involving pairs of $G_{21}$ (like those in Fig.~\ref{Figure-8}(c)),
with the composite structure acquiring the overall phase factor $e^{-i \mathbf{q} \cdot (\mathbf{r}_{2} + \mathbf{r}_{4} + \mathbf{r}_{1} + \mathbf{r}_{3})}$;
or else  pairs of $G_{12}$ (like those in Fig.~\ref{Figure-8}(d)), with the composite structure acquiring the overall phase factor 
$e^{i \mathbf{q} \cdot (\mathbf{r}_{2} + \mathbf{r}_{4} + \mathbf{r}_{1} + \mathbf{r}_{3})}$. 
To obtain eventually the self-energies $\Sigma_{i_{1}i_{2}}(x_{1},x_{2};\mathbf{q})$ within the $t$-matrix approximation, one has to add to each of these structures an upper single-particle line represented 
by the dashed line in each panel of Fig.~\ref{Figure-8},  which closes the fermion loop.
Owing again to the properties (\ref{G11-and-q-vector})-(\ref{G22-and-q-vector}), one ends up with the self-energies $\Sigma_{i_{1}i_{2}}(x_{1},x_{2};\mathbf{q})$ which themselves satisfies properties like 
(\ref{G11-and-q-vector})-(\ref{G22-and-q-vector}).
This proves our statement.

Finally, by combining the structures $\tilde{\Gamma}_{11}$ and $\tilde{\Gamma}_{22}$ in all possible ways  through  rungs of pairs of either $G_{12}$ or $G_{21}$ \cite{APS-2003}
( now  in the presence of a superfluid current), once the above phase factor $e^{\pm i \mathbf{q} \cdot (\mathbf{r}_{2} + \mathbf{r}_{4} \pm \mathbf{r}_{1} \pm \mathbf{r}_{3})}$ have conveniently been disposed off, one obtains  the following expressions  for the components $\Gamma_{ii'}(Q;\mathbf{q})$ of the full particle-particle ladder in   the broken-symmetry phase in the presence of a supercurrent:
\begin{eqnarray}
\left[ \! \begin{array}{cc} \Gamma_{11}(Q;\mathbf{q}) & \Gamma_{12}(Q;\mathbf{q}) \\ \Gamma_{21}(Q;\mathbf{q}) & \Gamma_{22}(Q;\mathbf{q}) \end{array} \! \right]
& = & \frac{1}{ A(Q;\mathbf{q}) \, A(-Q;\mathbf{q}) - B(Q;\mathbf{q})^{2}} 
\nonumber \\
& \times &  \left[ \! \begin{array}{cc} A(-Q;\mathbf{q}) & B(Q;\mathbf{q}) \\ B(Q;\mathbf{q}) & A(Q;\mathbf{q}) \end{array} \! \right] \, .
\label{full-particle-particle-ladder}
\end{eqnarray}
Here,
\begin{eqnarray}
A(Q;\mathbf{q}) & = & - \frac{m}{4 \pi a_{F}} + \int \! \frac{d \mathbf{k}}{(2 \pi)^{3}} \, \frac{m}{\mathbf{k}^{2}} 
\nonumber \\
& - & \sum_{k} \, \mathcal{G}_{11}^{\mathrm{mf}}(k + Q;\mathbf{q}) \, \mathcal{G}_{11}^{\mathrm{mf}}(-k;\mathbf{q})
\label{A} \\
B(Q;\mathbf{q}) & = & \sum_{k} \, \mathcal{G}_{12}^{\mathrm{mf}}(k + Q;\mathbf{q}) \, \mathcal{G}_{12}^{\mathrm{mf}}(-k;\mathbf{q})
\label{B}
\end{eqnarray}
where the regularization (\ref{regularization}) and the short-hand notation (\ref{notation-integral-sum-fermionic}) have been utilized.
 
Consistently with the non-self-consistent $t$-matrix approach in  the broken-symmetry phase considered in Ref.~\cite{PPS-2004}, the expressions for $\mathcal{G}_{11}^{\mathrm{mf}}(k;\mathbf{q})$ and $\mathcal{G}_{12}^{\mathrm{mf}}(k;\mathbf{q})$ entering  Eqs.~(\ref{A}) and (\ref{B}) are taken of the mean-field form (\ref{simplified-solutions-mean-field-11}) and 
(\ref{simplified-solutions-mean-field-12}), respectively.
By performing the sum over the fermionic Matsubara frequencies $\omega_{n}$ in the particle-particle bubbles of Eqs.~(\ref{A}) and (\ref{B}), we obtain eventually:
\begin{widetext}
\begin{eqnarray}
&& \sum_{k} \, \mathcal{G}_{11}^{\mathrm{mf}}(k + Q;\mathbf{q}) \, \mathcal{G}_{11}^{\mathrm{mf}}(-k;\mathbf{q}) = \int \! \frac{d \mathbf{k}}{(2 \pi)^{3}}
\label{diagonal-bubble} \\
& \times & \!\! \left\{ \!\! \frac{ u(\mathbf{k+Q};\mathbf{q})^{2} u(\mathbf{k};\mathbf{q})^{2} \left[ f_{F}(E_{+}(\mathbf{k+Q};\mathbf{q})) \! + \! f_{F}(E_{-}(\mathbf{k};\mathbf{q})) \! - \! 1 \right] } 
{ i \Omega_{\nu} - \frac{\mathbf{Q}\cdot\mathbf{q}}{m} - E(\mathbf{k+Q};\mathbf{q}) - E(\mathbf{k};\mathbf{q}) } \right.
+ \!\! \frac{ v(\mathbf{k+Q};\mathbf{q})^{2} v(\mathbf{k};\mathbf{q})^{2} \left[ 1 \! - \! f_{F}(E_{-}(\mathbf{k+Q};\mathbf{q})) \! - \! f_{F}(E_{+}(\mathbf{k};\mathbf{q})) \right] } 
{ i \Omega_{\nu} - \frac{\mathbf{Q}\cdot\mathbf{q}}{m} + E(\mathbf{k+Q};\mathbf{q}) + E(\mathbf{k};\mathbf{q}) }
\nonumber \\
& + & \,  \left. \frac{ u(\mathbf{k+Q};\mathbf{q})^{2} v(\mathbf{k};\mathbf{q})^{2} \left[ f_{F}(E_{+}(\mathbf{k+Q};\mathbf{q})) \! - \! f_{F}(E_{+}(\mathbf{k};\mathbf{q})) \right] } 
{ i \Omega_{\nu} - \frac{\mathbf{Q}\cdot\mathbf{q}}{m} - E(\mathbf{k+Q};\mathbf{q}) + E(\mathbf{k};\mathbf{q}) } 
+ \!\! \frac{ v(\mathbf{k+Q};\mathbf{q})^{2} u(\mathbf{k};\mathbf{q})^{2} \left[ f_{F}(E_{-}(\mathbf{k};\mathbf{q})) \! - \! f_{F}(E_{-}(\mathbf{k+Q};\mathbf{q})) \right] } 
{ i \Omega_{\nu} - \frac{\mathbf{Q}\cdot\mathbf{q}}{m} + E(\mathbf{k+Q};\mathbf{q}) - E(\mathbf{k};\mathbf{q}) }   \right\}
\nonumber
\end{eqnarray}
and
\begin{eqnarray}
&& \sum_{k} \, \mathcal{G}_{12}^{\mathrm{mf}}(k + Q;\mathbf{q}) \, \mathcal{G}_{12}^{\mathrm{mf}}(-k;\mathbf{q}) = \int \! \frac{d \mathbf{k}}{(2 \pi)^{3}}
u(\mathbf{k+Q};\mathbf{q}) \, v(\mathbf{k+Q};\mathbf{q}) \, u(\mathbf{k};\mathbf{q}) \, v(\mathbf{k};\mathbf{q})
\nonumber \\
& \times & \,\, \left\{ \frac{ \left[ f_{F}(E_{+}(\mathbf{k+Q};\mathbf{q})) \! + \! f_{F}(E_{-}(\mathbf{k};\mathbf{q})) \! - \! 1 \right] } 
{ i \Omega_{\nu} - \frac{\mathbf{Q}\cdot\mathbf{q}}{m} - E(\mathbf{k+Q};\mathbf{q}) - E(\mathbf{k};\mathbf{q}) } \right.
+ \,\,\, \frac{ \left[ 1 \! - \! f_{F}(E_{-}(\mathbf{k+Q};\mathbf{q})) \! - \! f_{F}(E_{+}(\mathbf{k};\mathbf{q})) \right] } 
{ i \Omega_{\nu} - \frac{\mathbf{Q}\cdot\mathbf{q}}{m} + E(\mathbf{k+Q};\mathbf{q}) + E(\mathbf{k};\mathbf{q}) }
\nonumber \\
& - & \left. \,\,\,\, \frac{ \left[ f_{F}(E_{+}(\mathbf{k+Q};\mathbf{q})) \! - \! f_{F}(E_{+}(\mathbf{k};\mathbf{q})) \right] } 
{ i \Omega_{\nu} - \frac{\mathbf{Q}\cdot\mathbf{q}}{m} - E(\mathbf{k+Q};\mathbf{q}) + E(\mathbf{k};\mathbf{q}) } 
-  \,\,\, \frac{ \left[ f_{F}(E_{-}(\mathbf{k};\mathbf{q})) \! - \! f_{F}(E_{-}(\mathbf{k+Q};\mathbf{q})) \right] } 
{ i \Omega_{\nu} - \frac{\mathbf{Q}\cdot\mathbf{q}}{m} + E(\mathbf{k+Q};\mathbf{q}) - E(\mathbf{k};\mathbf{q}) } \right\} \, .
\label{off-diagonal-bubble}
\end{eqnarray}
\end{widetext}
These expressions coincide with those reported in Ref.~\cite{Taylor-2006} for the same particle-particle bubbles in the presence of a superfluid flow.
Note that everywhere in the above expressions the bosonic Matsubara frequency $i\Omega_{\nu}$ is ``Doppler shifted'' by the amount $- \frac{\mathbf{Q}\cdot\mathbf{q}}{m}$.
Note also that the gapless condition of the particle-particle ladder (\ref{full-particle-particle-ladder}) at  $Q=0$ is guaranteed by enforcing the gap equation in the presence of a superfluid flow, 
in the form:
\begin{eqnarray}
&& B(Q=0;\mathbf{q}) - A(Q=0;\mathbf{q}) = \frac{m}{4 \pi a_{F}}
\nonumber \\
& + & \int \! \frac{d \mathbf{k}}{(2 \pi)^{3}} \!
\left\{ \frac{ \left[ 1 - 2 f_{F}(E_{+}(\mathbf{k};\mathbf{q})) \right] } { 2 E(\mathbf{k};\mathbf{q}) } - \frac{m}{\mathbf{k}^{2}} \right\} = 0 \, .
\label{gap-equation-with_current}
\end{eqnarray}
It should be remarked that this expression coincides with the LPDA equation in the presence of a super-current of Sec.~\ref{sec:numerical_procedures}, in the limiting case of a homogeneous system.

\section{MAPPING ONTO THE BOSONIC TWO-FLUID MODEL IN THE BEC LIMIT \\ OF THE BCS-BEC CROSSOVER}
\label{sec:Appendix-B}

In this Appendix, we consider the strong-coupling (BEC) limit of the expression (\ref{G-11-pairing-flucuations_and_current}) for the diagonal (normal) single-particle Green's function $\mathcal{G}_{11}^{\mathrm{pf}}(k;\mathbf{q})$ which includes pairing fluctuations in the presence of a superfluid flow at the level of the $t$-matrix approximation, in terms of which we will obtain the expressions of the number and current densities in this limit.

In the following, we shall consider the homogeneous case, whereby the expressions  (\ref{local-density-pairing-fluctuations}) for the number density and (\ref{local-current-pairing-fluctuations}) for the current density reduce to:
\begin{equation}
n = 2 \! \int \!\!\! \frac{d \mathbf{k}}{(2 \pi)^{3}} \, \frac{1}{\beta} \sum_{n} e^{i \omega_{n} \eta} \, \mathcal{G}_{11}^{\mathrm{pf}}(\mathbf{k},\omega_{n};\mathbf{q})
\label{density-pairing-fluctuations}
\end{equation}
and
\begin{equation}
\mathbf{j} = \frac{\mathbf{q}}{m} \, n +  2 \! \int \!\!\! \frac{d \mathbf{k}}{(2 \pi)^{3}} \frac{\mathbf{k}}{m} \, \frac{1}{\beta} \sum_{n} e^{i \omega_{n} \eta} \, \mathcal{G}_{11}^{\mathrm{pf}}(\mathbf{k},\omega_{n};\mathbf{q}) \, .
\label{current-pairing-fluctuations}
\end{equation}
We will show that, in the BEC limit of the BCS-BEC crossover, these expressions are consistent with a \emph{bosonic two-fluid model} for the composite bosons (or dimers) 
that form in this limit, specifically at the level of the Bogoliubov approximation.

To this end, we begin by considering the expressions (\ref{A}) and (\ref{diagonal-bubble}) for $A(Q;\mathbf{q})$ and (\ref{B}) and (\ref{off-diagonal-bubble}) for $B(Q;\mathbf{q})$.
In the BEC limit, $|\mu|$ is the largest energy scale in the problem such that we can consider the limit $\beta \mu \rightarrow - \infty$, with the residual term $\mu_{B} = 2 \mu + (m a_{F})^{-1}$ corresponding to the dimers chemical
potential \cite{Physics-Reports-2018}.
Accordingly, in the expressions (\ref{diagonal-bubble}) and (\ref{off-diagonal-bubble}) we can neglect all Fermi functions, but we have to keep the wave vector $\mathbf{q}$ in the denominators therein, 
both in the expressions of $E(\mathbf{k};\mathbf{q})$ and $E(\mathbf{k+Q};\mathbf{q})$ where it occurs in the combination $- \mu + \frac{\mathbf{q}^{2}}{2m} = \frac{1}{2 m a_{F}^{2}} - \frac{1}{2} \bar{\mu}_{B}$ with
\begin{equation}
\bar{\mu}_{B} \equiv \mu_{B} - \frac{(2 \mathbf{q})^{2}}{4m} \, ,
\label{bosonic-chemical-potential}
\end{equation}
as well as in the Doppler shifted frequency $i \Omega_{\nu} - \frac{\mathbf{Q}\cdot\mathbf{q}}{m}$.
In addition, the expressions (\ref{diagonal-bubble}) and (\ref{off-diagonal-bubble}) can be expanded to the leading order in the dimers kinetic energy $\frac{\mathbf{Q}^{2}}{4m}$, since on physical grounds 
this may  be comparable with  the bosonic chemical potential $\mu_{B}$.

We can then resort to the analytic methods of Ref.~\cite{MPS-1998} and obtain in the BEC limit (cf. also Appendix A of Ref.~\cite{APS-2003})
\begin{eqnarray}
A(Q;\mathbf{q}) & \simeq & \frac{m^{2} a_{F}}{8 \pi} \! \left[ \bar{\mu}_{B} + \frac{\mathbf{Q}^{2}}{4m} - \left( i \Omega_{\nu} - \frac{\mathbf{Q}\cdot\mathbf{q}}{m} \right) \right]
\label{A-BEC-limit} \\
B(Q;\mathbf{q}) & \simeq & \frac{m^{2} a_{F}}{8 \pi} \, \bar{\mu}_{B} \, ,
\label{B-BEC-limit}
\end{eqnarray}
such that from Eq.~(\ref{full-particle-particle-ladder})
\begin{equation}
\Gamma_{11}(Q;\mathbf{q}) \simeq \frac{8 \pi}{m^{2} a_{F}} \, \frac{ \left[ \bar{\mu}_{B} + \frac{\mathbf{Q}^{2}}{4m} + \left( i \Omega_{\nu} - \frac{\mathbf{Q}\cdot\mathbf{q}}{m} \right) \right] }
                                                                                                          {E_{B}(\mathbf{Q})^{2} - \left( i \Omega_{\nu} - \frac{\mathbf{Q}\cdot\mathbf{q}}{m} \right)^{2} }
\label{11-particle-particle-ladder-BEC-limit}
\end{equation}
where
\begin{equation}
E_{B}(\mathbf{Q}) = \sqrt{ \left( \bar{\mu}_{B} + \frac{\mathbf{Q}^{2}}{4m} \right)^{2} - \bar{\mu}_{B}^{2} }
\label{Bogoliubov-dispersion}
\end{equation}
is the Bogoliubov dispersion relation 	with $\bar{\mu}_{B} = \frac{2 k_{F}^{3} a_{F}}{3 \pi m} = \frac{4 \pi (2 a_{F})}{(2m)} \! \left(\frac{n}{2}\right)$.
This expression corresponds to the value $a_{B} = 2 a_{F}$ of the dimer-dimer scattering length, consistently with that  obtained in the BEC limit of the LPDA equation \cite{SS-2014} when it reduces to the GP equation (as, in turn, obtained from the BdG equations in that limit \cite{PS-2003}).
The expression (\ref{11-particle-particle-ladder-BEC-limit}) can further be cast in the meaningful form
\begin{equation}
\Gamma_{11}(Q;\mathbf{q}) \simeq - \frac{8 \pi}{m^{2} a_{F}} \! \left[ \! \frac{ u_{B}(\mathbf{Q})^{2} } { i \Omega_{\nu} - E^{B}_{+}(\mathbf{Q};\mathbf{q}) } \! - \! 
                                                                                                                 \frac{ v_{B}(\mathbf{Q})^{2} } { i \Omega_{\nu} + E^{B}_{-}(\mathbf{Q};\mathbf{q}) } \! \right]
\label{11-particle-particle-ladder-BEC-limit-meaningful}
\end{equation}
where $E^{B}_{\pm}(\mathbf{Q};\mathbf{q}) = E_{B}(\mathbf{Q}) \pm \frac{\mathbf{Q}\cdot\mathbf{q}}{m}$ and
\begin{eqnarray}
u_{B}(\mathbf{Q})^{2} & = & \frac{1}{2} \! \left( \frac{\bar{\mu}_{B} + \frac{\mathbf{Q}^{2}}{4m} }{E_{B}(\mathbf{Q})} + 1 \right)
\label{Bogoliubov-u} \\
v_{B}(\mathbf{Q})^{2} & = & \frac{1}{2} \! \left( \frac{\bar{\mu}_{B} + \frac{\mathbf{Q}^{2}}{4m} }{E_{B}(\mathbf{Q})} - 1 \right)
\label{Bogoliubov-v}
\end{eqnarray}
are the Bogoliubov coherence factors  such that $u_{B}(\mathbf{Q})^{2} - v_{B}(\mathbf{Q})^{2} = 1$.

The result (\ref{11-particle-particle-ladder-BEC-limit-meaningful}) can now be used in the expression (\ref{self-energies-t-matrix}) for the diagonal fermionic self-energy $\mathfrak{S}_{11}^{\mathrm{pf}}(k;\mathbf{q})$, 
which in the BEC limit can be further approximated by
\begin{equation}
\mathfrak{S}_{11}^{\mathrm{pf}}(k;\mathbf{q}) \! \simeq \! - \sum_{Q} \Gamma_{11}(Q;\mathbf{q}) \, \mathcal{G}^{(0)}(Q \! - \! k;\mathbf{q}) 
\label{self-energies-t-matrix-BEC-limit}
\end{equation}
where $\mathcal{G}^{(0)}(k;\mathbf{q}) = [ i\omega_{n} - \xi(\mathbf{k+q}) ]^{-1}$.
By a similar token, the diagonal single-particle Green's function $\mathcal{G}_{11}^{\mathrm{pf}}(k;\mathbf{q})$ to be used in Eqs.~(\ref{density-pairing-fluctuations}) and (\ref{current-pairing-fluctuations}) takes the form
\begin{eqnarray}
& & \mathcal{G}_{11}^{\mathrm{pf}}(k;\mathbf{q}) \simeq \mathcal{G}^{(0)}(k;\mathbf{q}) + \mathcal{G}^{(0)}(k;\mathbf{q}) \, \mathfrak{S}_{11}^{\mathrm{pf}}(k;\mathbf{q}) \, \mathcal{G}^{(0)}(k;\mathbf{q})
\nonumber \\
& + & \mathcal{G}^{(0)}(k;\mathbf{q}) \, (\! - \Delta_{\mathbf{q}}) (- \mathcal{G}^{(0)}(-k;\mathbf{q})) (\! - \Delta_{\mathbf{q}}) \mathcal{G}^{(0)}(k;\mathbf{q}).
\label{Greens-function-BEC-limit}
\end{eqnarray}
We are thus left with entering the results (\ref{11-particle-particle-ladder-BEC-limit-meaningful}), (\ref{self-energies-t-matrix-BEC-limit}), and (\ref{Greens-function-BEC-limit}) into the expressions (\ref{density-pairing-fluctuations}) for the density and (\ref{current-pairing-fluctuations}) for the current.

In both cases, we make use of the following result
\begin{eqnarray}
& \frac{1}{\beta} \sum_{n} & \! e^{i \omega_{n} \eta} \! \left( \! \frac{1}{i \omega_{n} - \xi(\mathbf{k+q})} \! \right)^{2} \! \frac{1}{i \Omega_{\nu} - i \omega_{n} - \xi(\mathbf{Q-k+q})}
\nonumber \\
& \simeq & \!\!\! - \, \frac{1}{ \left[ \xi(\mathbf{k+q}) + \xi(\mathbf{Q-k+q}) - i \Omega_{\nu} \right]^{2}}
\nonumber \\
& \simeq & \!\!\! - \, \frac{1}{4 \, \xi(\mathbf{k})^{2}} \left[ 1 + \frac{\mathbf{k}\cdot\mathbf{Q}/m}{\xi(\mathbf{k})} \right] \, ,
\label{useful-result}
\end{eqnarray}
where in the second line we have exploited the BEC limit and in the third line we have retained the leading significant terms in $\mathbf{Q}$.
In particular, in the third line of Eq.~(\ref{useful-result}) the first term within brackets contributes to the expression (\ref{density-pairing-fluctuations}) of the density 
while the second term within brackets contributes to the expression (\ref{current-pairing-fluctuations}) of the current.

In the BEC limit, the expression (\ref{density-pairing-fluctuations}) of the density thus becomes:
\begin{eqnarray}
n & \simeq & 2 \left( \Delta_{\mathbf{q}}^{2} + \sum_{Q} e^{i \Omega_{\nu} \eta} \, \Gamma_{11}(Q;\mathbf{q}) \right) \!\! \int \!\!\! \frac{d \mathbf{k}}{(2 \pi)^{3}} \, \frac{1}{4 \, \xi(\mathbf{k})^{2}} 
\nonumber \\
& = &  \Delta_{\mathbf{q}}^{2} \, \frac{m^{2} a_{F}}{4 \pi} \, - 2 \! \int \!\!\! \frac{d \mathbf{Q}}{(2 \pi)^{3}} \, \frac{1}{\beta} \sum_{\nu} e^{i \Omega_{\nu} \eta}
\nonumber \\
& \times & \left[ \! \frac{ u_{B}(\mathbf{Q})^{2} } { i \Omega_{\nu} - E^{B}_{+}(\mathbf{Q};\mathbf{q}) } \! - \! \frac{ v_{B}(\mathbf{Q})^{2} } { i \Omega_{\nu} + E^{B}_{-}(\mathbf{Q};\mathbf{q}) } \! \right]
\nonumber \\
& = & n_{0} + 2 \! \int \!\!\! \frac{d \mathbf{Q}}{(2 \pi)^{3}} \! \left[ v_{B}(\mathbf{Q})^{2} + \left( u_{B}(\mathbf{Q})^{2} + v_{B}(\mathbf{Q})^{2} \right) \right.
\nonumber \\
& & \hspace{3.7cm}  \left. \times \, b(E^{B}_{+}(\mathbf{Q};\mathbf{q})) \right]
\label{density-Bogoliubov}
\end{eqnarray}
where $n_{0} = \Delta_{\mathbf{q}}^{2} \, (m^{2} a_{F})/(4 \pi) $ is (twice the value of) the condensate density of composite bosons that form in this limit \cite{PS-2003} and $b(\epsilon) = (e^{\beta \epsilon} - 1)^{-1}$ is the Bose function.
This result is what would be obtained for a bosonic gas treated with the Bogoliubov approximation in the presence of a supercurrent.
It represents the bosonic counterpart of the fermionic result (\ref{local-density-mean-field}), once specified to the homogeneous case.

In a related fashion, the expression (\ref{current-pairing-fluctuations}) of the current becomes:
\begin{eqnarray}
\mathbf{j} & - & \frac{\mathbf{q}}{m} \, n \simeq 2 \sum_{Q} e^{i \Omega_{\nu} \eta} \, \Gamma_{11}(Q;\mathbf{q}) \!\! \int \!\!\! \frac{d \mathbf{k}}{(2 \pi)^{3}} \, \frac{\mathbf{k}}{4 \, m \, \xi(\mathbf{k})^{2}} \, 
                                                                                                                                                                                                     \frac{\mathbf{k}\cdot\mathbf{Q}/m}{\xi(\mathbf{k})} 
\nonumber \\
& = & 2 \sum_{Q} e^{i \Omega_{\nu} \eta} \, \mathbf{Q} \, \Gamma_{11}(Q;\mathbf{q}) \!\! \int \!\!\! \frac{d \mathbf{k}}{(2 \pi)^{3}} \, \frac{\mathbf{k}^{2}}{12 \, m \, \xi(\mathbf{k})^{3}} 
\nonumber \\
& = & 2  \left( \frac{m a_{F}}{16 \pi} \right) \, \sum_{Q} e^{i \Omega_{\nu} \eta} \, \mathbf{Q} \, \Gamma_{11}(Q;\mathbf{q})
\nonumber \\
& = & - 2 \sum_{Q} e^{i \Omega_{\nu} \eta} \, \frac{\mathbf{Q}}{2m} \! 
                                             \left[ \! \frac{ u_{B}(\mathbf{Q})^{2} } { i \Omega_{\nu} - E^{B}_{+}(\mathbf{Q};\mathbf{q}) } \! + \! \frac{ v_{B}(\mathbf{Q})^{2} } { i \Omega_{\nu} + E^{B}_{+}(\mathbf{Q};\mathbf{q}) } \! \right]
\nonumber \\
& = & 2 \! \int \!\!\! \frac{d \mathbf{Q}}{(2 \pi)^{3}} \, \frac{\mathbf{Q}}{2m} \,  b(E^{B}_{+}(\mathbf{Q};\mathbf{q})) \, .
\label{current-Bogoliubov} 
\end{eqnarray}

This expression has the typical form of the current within a two-fluid model \cite{Pethick-Smith-2008} for a bosonic gas treated with the Bogoliubov approximation.
As such, it represents the bosonic counterpart of the fermionic result (\ref{local-current-mean-field}) once specified to the homogeneous case.
Note that the condensate density $n_{0}$ does not explicitly contribute to the thermal part of the current (\ref{current-Bogoliubov}), but it does only implicitly through the bosonic
chemical potential.




\begin{thebibliography}{99}
\bibitem{footnote-local-phase-transformation} The structure of Eqs.(\ref{G11-and-q-vector})-(\ref{Gij-Fourier-transform}) is consistent with the local phase transformation 
                                                                         $\Psi_{1}(\mathbf{r})=e^{i \mathbf{q}\cdot\mathbf{r}}\tilde{\Psi}_{1}(\mathbf{r})$ and $\Psi_{2}(\mathbf{r})=e^{-i \mathbf{q}\cdot\mathbf{r}}\tilde{\Psi}_{2}(\mathbf{r})$ 
                                                                         of the field operators in Nambu representation,
                                                                         which transforms from the lab frame where the superfluid is in motion to the rest frame of the superfluid where it is in equilibrium.
\end{thebibliography}
\end{document}